\documentstyle[german,pre,aps]{revtex}
\newenvironment{eqna}[1]{\begin{equation}\begin{array}{#1}}
                        {\end{array}\end{equation}}

\renewcommand{\vec}[1]{{\mathbf{#1}}}

\newcommand{\mat}[1]{{\mathbf{#1}}}

\newcommand{\dt}{{d \over dt}}

\draft
\begin{document}

\title{Hyperchaos in the generalized R"oss\-ler system}
\author{Th.~Meyer, M.J.~B"unner, A.~Kittel, J.~Parisi}
\address{Faculty of Physics, 
         Department of Energy and Semiconductor Research, 
         University of Oldenburg, D-26111 Oldenburg, Germany}
\date{\today}
\maketitle
\begin{abstract}
Introduced as a model for hyperchaos, the generalized R"oss\-ler 
system of dimension $N$ is obtained by linearly coupling $N-3$ 
additional degrees of freedom to the original R"oss\-ler equation. 
Under variation of a single control parameter, it is able to 
exhibit the chaotic hierarchy ranging from fixed points via limit 
cycles and tori to chaotic and, finally, hyperchaotic attractors. 
By the help of a mode transformation, we reveal a structural 
symmetry of the generalized R"oss\-ler system. The latter will 
allow us to interpret number, shape, and location in phase space of 
the observed coexisting attractors within a common scheme for 
arbitrary odd dimension $N$. The appearance of hyperchaos is 
explained in terms of interacting coexisting attractors. In a 
second part, we investigate the Lyapunov spectra and related 
properties of the generalized R"oss\-ler system as a function of 
the dimension $N$. We find scaling properties which are not similar 
to those found in homogeneous, spatially extended systems, 
indicating that the high-dimensional chaotic dynamics of the 
generalized R"ossler system fundamentally differs from 
spatio-temporal chaos. If the time scale is chosen properly, 
though, a universal scaling function of the Lyapunov exponents is 
found, which is related to the real part of the eigenvalues of an 
unstable fixed point.
\end{abstract}

\section{introduction}

Chaotic dynamics has been intensively investigated with the help of 
simple low-dimensional models like the Lorenz \cite{Lor63} or the 
R"oss\-ler system \cite{Roe76}. Because of the restricted phase 
space, only low-dimensional chaotic motion is observed in these 
systems. Low-dimensional chaos can also be observed in nature 
\cite{jfh95}, although the underlying dynamical systems have an 
infinite number of degrees of freedom. One may wonder: What is the 
connection of this low-dimensional chaos to the truly 
high-dimensional states, that may arise out of chaos under 
variation of one control parameter? How does the transition from
the low-dimensional to the high-dimensional states take place?

Consider the attractors of dissipative ordinary differential 
equations under variation of the dimension $N$ of phase space. 
R"ossler~\cite{Roe83} has postulated a chaotic hierarchy where more 
and more qualitatively new forms of complex motion develop with 
increasing dimension: In one dimension, only stable fixed points 
are encountered. In two dimensions, also periodic orbits can exist. 
In three dimensions, the possibility of quasiperiodicity and chaos 
arises. Thus, at the lower end of the hierarchy, the well-known 
low-dimensional dynamical states of motion can be found. Somewhere 
high up in the hierarchy, things like turbulence or noise may be 
located. One crucial question is, how one can distinguish and 
characterize the higher steps of the hierarchy. A provisional 
classification can be made in terms of Lyapunov 
exponents~\cite{KB91b}. For dynamical states with more than one 
positive Lyapunov exponent, the term hyperchaos has been coined 
\cite{Roe79}. In order to study the full chaotic hierarchy, Baier 
and Sahle \cite{BS95} have introduced a class of model equations, 
the \emph{Generalized R"ossler System} (furtheron called 
GRS). Starting from the R"oss\-ler system as one of the simplest and
best understood nonlinear ordinary differential equations that 
exhibit chaos, the GRS is obtained by linearly coupling additional 
degrees of freedom to the original R"ossler system. The structure 
of the GRS of dimension $N$ is that of an $N-1$ dimensional linear 
subsystem that is coupled to one nonlinear variable. As will be 
shown, the GRS preserves essential characteristics of the 
R"oss\-ler system while extending it to a phase space of arbitrary 
dimension $N$. This allows to study the influence of the discrete 
parameter $N$, without having to compare completely different 
systems.

Baier and Sahle~\cite{BS95} demonstrated that the GRS does, indeed, 
show hyperchaotic dynamics with an increasing number of positive 
Lyapunov exponents for increasing $N$. The GRS realizes one 
possible path through the complete chaotic hierarchy from a stable 
fixed point via periodic orbits and chaos up to hyperchaos. 
We~\cite{MBKP95} have introduced a mode transformation of the GRS 
based on the numerical solution of the linear subsystem. In the 
present paper, we will restate the mode transformation based on a 
semianalytical solution of the linear subsystem. The mode 
transformation will then be used to analyze the dynamics of the GRS 
with arbitrary dimension $N$ in phase space. In 
Section~\ref{chap1}, the GRS is introduced and general properties 
of the GRS are discussed. In Section~\ref{chap2}, we deal with the 
mode transformation and the concept of structural symmetry, that 
later on will allow us to understand the number, the form, and the 
location of the coexisting attractors of the GRS. Subsequently, we 
numerically investigate the dynamics of the GRS in phase space for 
the cases $N=5$ and $N=7$ in Section~\ref{chap3}. We interpret the 
observed dynamics within a general scheme that enables to predict 
the structure of the attractors for higher $N$. In 
Section~\ref{chap4}, then, we investigate the GRS in the 
hyperchaotic state for different values of the dimension $N$. We 
report scaling properties of the number of positive Lyapunov 
exponents, the Lyapunov dimension, the metric entropy, and the 
Lyapunov spectra 
as a function of $N$. The limit $N\longrightarrow\infty$ is 
discussed in view of the literature on that subject.

\section{General properties of the Generalized R"oss\-ler System}
\label{chap1}

The GRS is given by
\begin{eqnarray} 
     \dt \vec{x}(t) &=& \matrix A \vec x(t) - x_N(t) \vec e_{N-1}  
     \label{eq.GRS1}  
     \\
     \dt x_N(t)     &=& \varepsilon + bx_N(t) \left( x_{N-1}(t) -
d\right)
     \label{eq.GRS2}  
     \textrm{~,}
\end{eqnarray}
where
\[
A = 
    \left (
    \begin{array}{ccccc}
        a        &  -1        & 0        & \ldots   & 0        \\
        1        &   0        & -1       & \ddots   & \vdots   \\
        0        &   1        & \ddots   & -1       & 0        \\
        \vdots   &   \ddots   & 1        & 0        & -1       \\
        0        &   \ldots   & 0        & 1        & 0        
    \end{array}
    \right ) \textrm{, } 
    \vec x(t) = \left(\begin{array}{c}
                        x_1(t)    \\
                        \vdots \\
                        x_{N-1}(t)
                     \end{array}\right) \quad\textrm{, }
\]
\[
    \vec e_{N-1} = \left(\begin{array}{c}
                        0      \\
                        \vdots \\
                        0      \\
                        1      
                     \end{array}\right ) \textrm{, }
    \vec x(t),\vec e_{N-1} \in \mathcal{R}^{N-1} \textrm{~.}
\]
It is composed of an $N-1$ dimensional linear subsystem $\vec x$ 
and one nonlinear variable $x_N$. Qualitatively, the mechanism of 
instability of the GRS is the same as that of the R"oss\-ler system 
for all dimensions $N$. The positive feedback or autocatalytic 
process that is controlled by the parameter $a$ causes an expansive 
dynamics of the GRS around the origin. As long as $x_{N-1}$ remains 
well below the threshold $d$ of $x_N$, $x_N$ adiabatically follows 
its equilibrium value $\varepsilon / b(d-x_{N-1})$ and does not 
influence the linear subsystem $\vec x$ appreciably. When $x_{N-1}$ 
comes close to or exceeds $d$, $x_N$ will start to grow rapidly, 
thereby folding the system back to a state of lower amplitude (via 
$\vec e_{N-1}$ in Eq.~(\ref{eq.GRS1})). This time development of 
$x_N$ of long intervals of small amplitude interrupted by short 
spikes leads us to call $x_N$ the \textit{nonlinear trigger}.

One of the three parameters $\varepsilon$, $b$, and $d$ can be
eliminated by rescaling the amplitude of $(\vec x, x_N)$. In the 
form the equations are given in (\ref{eq.GRS1}), (\ref{eq.GRS2}), 
they are scaled in time in such a way that the (angular) 
eigenfrequencies of the linear subsystem lie in the interval [0,2] 
for all values of $N$ (see Section~\ref{chap2}).

The divergence of the GRS is $a + b(x_{N-1}-d)$, independent of $N$.
Thus, the GRS is dissipative, if $\left<x_{N-1}\right> < d-{a \over 
b}$ (the angle brackets denote the time average). This is true for 
all parameter values to be considered in the following.

\label{evenodd}
For odd $N$, the linear subsystem $\vec x$ can be transformed into 
$N-1\over 2$ harmonic oscillators that are coupled only via the
nonlinear trigger $x_N$, as will be shown in Section~\ref{chap2} 
(compare also \cite{MBKP95}). For even $N$, the linear subsystem 
can be transformed into $N-2\over 2$ oscillators together with one 
variable, which simply grows exponentially. That means, there is 
one positive real eigenvalue in addition to $N-2\over 2$ pairs of 
complex conjugate eigenvalues. This leads to a qualitatively 
different dynamics. See the remark at the end of 
subsection~\ref{modetrafo} for clarifying this point. In the present
paper, we restrict ourselves to the case of odd $N$.

\section{The Mode Picture}
\label{chap2}
In the following section, we develop the mode picture of the GRS. 
First, we will present the solution of the linear subsystem. 
Herewith, we can transform the GRS into a mode picture, where the 
linear subsystem $\vec x$ consists of harmonic oscillators, which 
are coupled only via the nonlinear trigger variable $x_N$. In the 
mode picture, we are in the position to explain the idea of the 
structural symmetry, which will be of central importance for 
understanding the dynamics of the GRS in subsequent sections. 
Finally, we show the fixed points together with their stability 
properties.

\subsection{Semianalytical solution of the linear subsystem}
\label{solutionlinsub}

Consider the linear subsystem $\vec x$ (Eq.~(\ref{eq.GRS1})). Here, 
we restate the equations in component form as a linear chain with 
boundary conditions:
\begin{eqna}{rcl}
    \label{eq.linsub}
    \dt x_n &=& x_{n-1} - x_{n+1} 
             \textrm{, } n \in \{1, \ldots, N-1\} \textrm{,}\\
     x_0 &=& a x_1 \textrm{ ,} \\
     x_N &=& 0 \textrm{ .}     \\ 
\end{eqna}
To solve the linear subsystem $\vec x$, an exponential ansatz is 
used:
\begin{eqna}{rcl}
    \label{eq.ansatz}
    x_{2n+1} &=&  \cos((2n+1)k+\varphi) e^{i \Omega(k) t} \\
    && \hspace{2cm}{
        \textrm{, for $n \in \left\{0,\ldots,{N-3 \over 2}\right\}$
,}}\\
    x_{2n}   &=& c\sin(2nk+\varphi)     e^{i \Omega(k) t} \\
    && \hspace{2cm}{ 
        \textrm{, for $n \in \left\{0,\ldots,{N-1 \over 2}\right\}$}
        \textrm{~,}}
\end{eqna}
where $k$ is the wave number of the eigenmodes, $\Omega(k)$ the 
corresponding frequency with a specific dispersion relation.
Substituting ansatz (\ref{eq.ansatz}) into chain (\ref{eq.linsub}) 
yields
\begin{eqna}{rcl}
    \label{frequenzen}
    \ c &=& -i \textrm{ ,}\\
    \Omega_{\pm m} &=& 2 \sin(k_{\pm m})
           \textrm{ , }m \in \{1,\ldots,{N-1\over 2}\}\textrm{ ,} \\
    k_{\pm m} &=& \pm {2m - 1 \over 2N} \pi - {1\over N} 
\varphi_{\pm m}
           \textrm{ ,}\\
    -i\sin(\varphi_{\pm m})
       &=& a \cos(\pm {2m - 1\over 2N}\pi+{N-1\over N}\varphi_{\pm 
m}) 
           \textrm{~.}\\
\end{eqna}
These $(N-1)$ eigenmodes are a complete solution of the $(N-1)$
dimensional linear subsystem $\vec x$ for odd $N$. The solution is 
analytical up to a complex correction $\varphi_{\pm m}$ to the 
allowed wave numbers $k_{\pm m}$, which is determined by a 
transcendent equation. For $a=0$, this equation can be solved giving
\begin{eqna}{lrcl}
    & \varphi &=& 0 \textrm{ ,}\\
    & k &=& {2m-1\over 2N} \pi 
            \textrm{ , where } m \in \{1,\ldots, {N-1\over 2}\}
\textrm{~.} 
\end{eqna}
For $a < 1$, there are $N-1\over 2$ pairs of complex conjugate
eigenvalues $(\Omega_m, \Omega_{-m})$, where $\Omega_m = 
-\bar\Omega_{-m}$. The corresponding pairs of eigenmodes represent 
harmonic oscillators with angular frequencies 
$\omega_m=\mathrm{\,Re\,} \Omega_m$ and autocatalytic coefficients 
$\alpha_m = -2 * \mathrm{\,Im\,} \Omega_m$. The complex corrections 
$\phi_{\pm m}$ to the values of $k_m$ for $a \ne 0$ can easily be 
expanded about $a=0$. The expansion is up to the order of $a^2$
\begin{eqna}{rcl}
    \phi_{\pm m} &=& i a\cos({2m-1\over 2N}\pi)  \\
                 &&  + {N-1\over 2N} a^2 \sin(\pm {2m-1\over N}\pi) 
                     + {\mathrm{O}} (a^3) \textrm{ .}\\
\end{eqna}
In the considered range of $a \in [0,0.35]$, this is an excellent 
approximation (however, in the subsequent analysis, we have 
numerically calculated the eigenvalues and eigenvectors of $\mat A$ 
in order to minimize errors). The imaginary part of $\varphi_m$ 
determines the autocatalytic coefficient $\alpha_m$. It is linear 
in $a$ up to ${\mathrm{O}}(a^3)$. The real part leads to a 
correction of $\omega_m$. But for $a \in [0,0.35]$, this correction 
is very small, i.e., the frequencies of the oscillators are almost 
independent of $a$ in the considered range. From 
equation~(\ref{frequenzen}), we infer that the frequencies of the 
oscillators are bounded ($0 < \omega < 2$ for all $N$) and we expect 
the autocorrelation time $\tau_c$ in the case of an unstable 
dynamics to be independent of $N$. Additionally, the velocity of 
signals of frequency $\omega$ traversing the linear subsystem 
equals $2\cos k$. Therefore, the time $\tau_s$ a signal takes 
to traverse the linear subsystem (from $x_1$ to $x_{(N-1)}$ or 
back, since the linear subsystem allows for a bidirectional 
transport of signals) is expected to be approximately $N/2$.

\subsection{The mode transformation}
\label{modetrafo}
We have solved the linear subsystem $\vec{x}$. The next step is to 
transform it into its eigensystem $\vec y$. The transformed 
equations for $\vec y$ are completely decoupled (as the matrix 
$\mat A$ is transformed into a diagonal matrix). Thus, we can treat 
the different oscillators independently. The idea of the following 
steps of the transformation is to bring the oscillators that 
correspond to pairs of complex conjugate eigenvalues into a form 
that is as similar to the oscillator of the original R"oss\-ler 
system as possible. The oscillators are of the form
\[
    \dt {y_{2m-1}(t) \choose y_{2m}(t)} = \left(\begin{array}{cc}
        i\Omega_m & 0 \\ 0 & -i\Omega_m \end{array}
    \right) {y_{2m-1}(t) \choose y_{2m}(t)} \textrm{~.}
\]
It can be considered as the principal value decomposition of the 
following R"ossler-like oscillator:
\[
    \dt {z_{2m-1}(t) \choose z_{2m}(t)}= \left(\begin{array}{cc}
        -\alpha_m & -\omega_m^2 \\ 1 & 0 \end{array}
    \right) {z_{2m-1}(t) \choose z_{2m}(t)} \textrm{~.}
\]
In the following, these oscillators are called \emph{oscillator 
$(z_{2m-1}, z_{2m})$}. For each variable of the linear subsystem 
$\vec x$, one can choose one complex factor of normalization. It 
can be utilized to make the entire transformation real (via phase 
factors) and to give the coupling to the nonlinearity $x_N$ -- 
which is not touched by the transformation -- a certain form (via 
amplitude factors). We choose this coupling in such a way that the 
resulting equations are
\begin{eqna}{rcl}
    \dt\vec{z}(t) &=& \mat B \vec{z}(t) - x_N(t) \vec c \\
    \dt x_N(t) &=& \varepsilon + bx_N(t)
        \left(\sum\limits_{m=1}^{{1\over2}(N-1)}z_{2m} - d\right) 
        \textrm{~,} 
\end{eqna}
where 
\[
\mat B = 
    \left (
    \begin{array}{ccccc}
        \alpha_1 & -\omega_1^2 &  0     & \cdots  & 0           \\
        1        & 0           &        & \ddots  & \vdots      \\ 
        0        &             & \ddots &         & 0           \\ 
        \vdots   & \ddots      &        & \alpha_{N-1\over2}
                                        & -\omega_{N-1\over2}^2 \\  
        0        & \cdots      & 0      & 1       & 0          
    \end{array}
    \right ) \textrm{~.}
\]
For the $\alpha_m$, it holds $\sum_{m=1}^{{1\over2}(N-1)} \alpha_m = 
a$, as the divergence of the GRS is not affected by the 
transformation. $\vec c$ denotes the vector of coupling constants 
of $x_N$ to the components of $\vec z$. If the complete 
transformation is called $\mat U$ (i.e. $\vec z = \mat U 
^{-1} \vec x$), then the vector of coupling constants $\vec c$ is
$\vec c = \mat U ^{-1} \vec e _{N-1}$. For $a=0$, we have $\vec c = 
{1\over N} {\left ( 0, (cos k_1)^2, 0,  (cos k_2)^2, \ldots, 0, 
(cos k_{N-1 \over 2})^2 \right)}^t$. This \emph{mode 
transformation} converts the GRS into a system of oscillators with 
frequencies $\omega_m$ and autocatalytic coefficients $\alpha_m$, 
that are coupled solely via the nonlinear trigger $x_N$. The 
coupling has a special form: the oscillators couple to the trigger 
only with the sum of their variables $z_{2m}$, i.e., they couple 
to $x_N$ identically. The coupling of $x_N$ back to $\vec z$ is
different for the different oscillators. In the following, we call 
the original form of the GRS the \textit{Baier-Sahle picture} and 
the transformed form the \textit{mode picture}. In 
Fig.~\ref{fi.trafo}, the structures of the Baier-Sahle picture and 
the mode picture are compared. Each linear degree of freedom is 
represented by a circle, each nonlinear degree of freedom by 
a square. Couplings are shown as connecting lines. In the 
Baier-Sahle picture, the similarity of the GRS to a spatially 
extended system is most obvious: the autocatalytic process and the 
nonlinear trigger $x_N$ can be considered as the boundary 
conditions of a homogeneous linear chain. On the other hand, in the 
mode picture, the dynamics of the GRS can be understood as the 
interaction of different oscillators, that are coupled only via one 
nonlinear trigger. As the trigger $x_N$ influences the dynamics of 
the linear subsystem only during the presence of the spikes, the 
effect of coupling is restricted to these short intervals in time. 
This allows to observe the dynamics of the different oscillators 
independently in the respective projections of phase space (see 
Section~\ref{chap3}).

For even values of $N$, one finds one positive real eigenvalue of 
the linear subsystem for $a>0$. In the mode picture, this 
corresponds to one exponentially growing degree of freedom, which 
is coupled to the nonlinear trigger $x_N$ in the same way as the 
even coordinates of the oscillators. The contribution of the 
nonlinear trigger to the time derivative of this exponential mode is
always negative. Thus, if once the exponential mode is pushed to a 
negative value by the nonlinear trigger it will escape to 
$-\infty$. This mechanism, in general, leads to a global instability 
of the GRS for even $N$.

\subsection{The structural symmetry}
The mode picture reveals the GRS to consist of harmonic oscillators 
$(z_{2m-1},z_{2m})$, that are identically coupled to the nonlinear
trigger $x_N$ (compare Fig.~\ref{fi.trafo}). A perfectly symmetric 
system of the same structure as that in Fig.~\ref{fi.trafo}(b), 
i.e., one with identical coefficients for each oscillator, would be 
symmetric with respect to any permutation of the oscillators. The 
invariant manifolds of such a system, specifically, the orbits and 
attractors, would have to obey this symmetry. For the invariant 
manifolds, there are two possibilities. Either they show the full 
symmetry in themselves, i.e., they are symmetric with respect to 
any permutation of the oscillators. Or they have coexisting mirror
images, such that the union of them obeys the full symmetry. In the 
GRS, the symmetry is broken merely by the difference in the 
frequencies $\omega_m$, autocatalytic coefficients $\alpha_m$, and 
coupling constants $c_n$. Nevertheless, the symmetry is preserved 
as a qualitative feature of the dynamics, as will be shown later 
on. We call this property \textit{structural symmetry}. For 
arbitrary odd $N$, the structural symmetry will be utilized to 
understand and predict number, shape, and location of the 
attractors of the GRS in phase space for moderate values of the 
autocatalytic coefficient $a$. The first manifestation of the 
structural symmetry can be seen in the stability properties of the 
fixed points of the GRS.

\subsection{The fixed points}

The fixed points of a dynamical system are the pivots around which 
the system evolves. Thus, it is essential to investigate the 
stability properties of the fixed points, if one wants to develop any
understanding of a dynamical system. For odd $N$, the GRS has two 
fixed points which are the same in all odd dimensions $N$, in the 
sense that the common linear coordinates and the nonlinear 
coordinate are identical for any two GRSs of different $N$. The two 
fixed points are (for odd $N$)%
\begin{eqna}{rl}
    x_1^{(1,2)} = x_3^{(1,2)} = \ldots = x_N^{(1,2)}
        & \\
    & \hspace{-1cm} = {d \over 2a} \mp
        \sqrt{\left({d \over 2a}\right)^2 - {\varepsilon \over ab}}
        \textrm{ ,}\\
    x_2^{(1,2)} = x_4^{(1,2)} = \ldots = x_{N-1}^{(1,2)} & \\
    & \hspace{-1cm} = {d \over 2} \mp
        \sqrt{\left({d \over 2}\right)^2 - {\varepsilon a \over b}}
        \textrm{\quad.}
\end{eqna}
Here, the minus signs belong to the index (1). The fixed point 
$(\vec x^{(1)}, x_N^{(1)})$ lies close to the origin of the system 
(for $\varepsilon=0$, it would be the origin). Accordingly, the 
stability analysis of this fixed point yields, in good 
approximation, 
the eigenmodes of the linear subsystem. The additional $N$-th 
eigenvalue is strongly attractive. It corresponds to the 
exponential decay of $x_N$ to $\varepsilon\over bd-x_{N-1}$ and its 
value is approximately $-bd$, as is to be expected. On the unstable 
manifold of this fixed point, the GRS expands until it is folded 
back by a trigger event of $x_N$. The presence of the nonlinear 
trigger $x_N$ slightly stabilizes the eigenmodes, such that they do 
no longer become unstable at $a=0$, but approximately at $a=0.006$. 
Interestingly, the oscillators that become unstable first are 
the ones that have a larger autocatalytic coefficient $\alpha_n$  
for higer values of $a$ (see Fig.~\ref{fi.fista}). 

At the fixed point $(\vec x^{(1)}, x_N^{(1)})$ , one can already
perceive the fingerprint of the structural symmetry: for each 
oscillator of the GRS, there is an oscillatory instability  of the 
inner fixed point $(\vec x^{(1)}, x_N^{(1)})$. The symmetry 
breaking is responsible for the difference in the eigenvalues of 
the fixed point. It will be shown in Section~\ref{chap3} that each 
of these instabilities of the fixed point $(\vec x^{(1)}, 
x_N^{(1)})$ gives rise to a coexisting attractor of the GRS. The 
fixed point $(\vec x^{(2)}, x_N^{(2)})$ governs the folding process
of the GRS. However, none of the attractors ever comes close to 
this fixed point.

\section{Dynamics of the Generalized R"oss\-ler System in phase 
space}
\label{chap3}

In the following section, the mode picture and the concept of 
structural 
symmetry will be utilized to develop an understanding of the 
dynamics of the GRS in the $N$-dimensional phase space. At first, 
we discuss the case $N=5$ in some detail. It will be demonstrated 
that for values of $a < 0.09$ two attractors coexist. For higher 
values of $a$, they interact in several crises. Finally, the two 
attractors merge into one hyperchaotic attractor. This scenario of 
three parameter regimes of coexisting attractors, interacting 
attractors, and one large hyperchaotic attractor will then be shown 
to hold in the case $N=7$ as well. Finally, we postulate that the 
dynamics of the GRS in arbitrary odd dimension $N$ can be explained 
by extrapolating this scheme. For all numerical calculations, we 
have restricted ourselves to the variation of $N$ and $a$. The other 
parameters remain constant at the values $\varepsilon=0.1$, $b=4$,
$d=2$.

\subsection{Numerical study of the case $N=5$}
In the mode picture, the five-dimensional GRS consists of two
oscillators that are coupled via the nonlinear trigger. First, we 
present a bifurcation diagram (Fig. \ref{fi.bif5}) and the 
corresponding Lyapunov spectrum (Fig. \ref{fi.ly5}) under variation 
of the control parameter $a$ (see Appendix for the numerical 
methods that were used). The bifurcation diagram consists of two 
separate parts. They correspond to two coexisting attractors and 
have been obtained using different initial conditions.

As one increases $a$ from $a=0$, the fixed point near the origin 
becomes unstable at about $a=0.0006$. Shortly one after the other, 
the two attractors emerge in two subsequent Hopf bifurcations as 
limit cycles. These two periodic orbits constitute the basis for 
all further development of the attractors. Although they become 
unstable at some point, they exist up to the highest values of $a$. 
In the following, they will be referred to as the \textit{period-1 
limit cycles} of the two attractors. \textit{Attractor 1} develops 
as a result of the instability that corresponds to oscillator 
$(z_1,z_2)$ in the mode picture. Accordingly, \textit{attractor 2} 
develops out of $(z_3,z_4)$. In Fig.~\ref{fi.orbit51}, two phase 
space projections of the period-1 limit cycle of attractor 1 are 
shown: one onto oscillator $(z_1,z_2)$ together with the nonlinear
trigger variable $x_5$ and the other onto oscillator $(z_3,z_4)$ 
togeter with $x_5$. The amplitude of oscillator $(z_3,z_4)$ is 
negligible compared with that of oscillator $(z_1,z_2)$. Here, the 
dynamics of the GRS is completely dominated by $(z_1,z_2)$ together 
with $x_5$. In general, the frequencies of the two oscillators are 
incommensurate. The two oscillators drift out of phase in each 
revolution. However, the spikes in $x_5$ cause phase shifts in 
$(z_3,z_4)$ that resynchronize the two oscillators. This leads to a 
1:2 mode locking. For higher $a$ (i.e., above $a=0.05$), oscillator 
$(z_1,z_2)$ can no longer force the phase locking and the periodic 
orbit breaks up into a quasiperiodic one. As one increases $a$ 
further, attractor 1 shows several periodic windows and, finally, 
low-dimensional chaos arises. In Fig.~\ref{fi.orbit52}, the phase 
space projections of this low-dimensional chaotic form of 
attractor~1 are depicted. Oscillator $(z_3,z_4)$ has still a much 
smaller amplitude than oscillator $(z_1,z_2)$ (though its influence 
on the nonlinear trigger $x_5$ is no longer negligible). And, 
remarkably, the dynamics in the three-dimensional projection onto 
$(z_1,z_2)$ together with $x_5$ shows a close resemblance to the 
R"oss\-ler system. It, therefore, can be considered as a 
'perturbed' R"ossler system.

The evolution of attractor 2 under variation of $a$ is completely 
analogous, except that now oscillator $(z_3,z_4)$ plays the dominant
role and $(z_1,z_2)$ is very small. In Fig.~\ref{fi.orbit53}, the 
period-1 limit cycle of attractor 2 is depicted. It corresponds to 
the periodic orbit of attractor 1 shown in Fig.~\ref{fi.orbit51}. 
Here, oscillator $(z_3,z_4)$ is dominant, whereas oscillator 
$(z_1,z_2)$ has a negligible amplitude. The phase locking between 
the oscillators is accomplished by phase shifts in oscillator 
$(z_1,z_2)$ due to the spikes in $x_5$ which are in turn triggered 
by $(z_3,z_4)$. Again, the dynamics in the projection, here onto 
$(z_3,z_4)$ together with $z_5$, resembles that of the R"oss\-ler 
system, which can best be seen on the low-dimensional chaotic orbit 
in Fig.~\ref{fi.orbit54} (compare the chaotic orbit on attractor 1 
in Fig.~\ref{fi.orbit52}).

The existence of the two attractors is a consequence of the 
structural symmetry: in a perfectly symmetric system (of the same 
structure as the GRS in the mode picture), each attractor, that is 
not in itself symmetric with respect to the exchange of any two 
oscillators (i.e., the transformation $z_1 \leftrightarrow z_3$, 
$z_2 \leftrightarrow z_4$ in the case $N=5$), must necessarily have 
one or more mirror images, such that their union fulfills the 
symmetry. An attractor, where oscillator $(z_1,z_2)$ dominates 
oscillator $(z_3,z_4)$, is obviously not symmetric with respect to 
the above transformation. Therefore, in the perfectly symmetric 
system, there would have to be another coexisting attractor, 
where oscillator $(z_3,z_4)$ dominates oscillator $(z_1,z_2)$. In the
GRS, the symmetry is broken by the difference in the frequencies,
autocatalytic constants, and coupling constants. However, as has 
been shown above, these two attractors do still exist. Moreover, 
their shape and location in phase space does, indeed, reflect the 
structural symmetry. Attractor 1 lies close to the subspace defined 
by $z_3 = 0$ and $z_4 = 0$. Attractor 2 is situated close to the 
subspace defined by $z_1=0$ and $z_2=0$. On attractor 1, oscillator 
$(z_1,z_2)$ is dominant and, on the periodic orbits, the phase 
locking is kept up by phase shifts of oscillator $(z_3,z_4)$. On 
attractor 2, accordingly, oscillator $(z_3,z_4)$ is dominant and, 
on the periodic orbits, the phase locking is kept up by phase 
shifts of oscillator $(z_1,z_2)$.

Up to $a\simeq 0.09$, the two attractors develop independently. 
They are located clearly separate in phase space. Then, a boundary 
crisis occurs: Attractor 1 collides with the separatrix to the 
basin of attractor 2 and ceases to exist as a stable dynamical 
state. It still attracts the trajectories out of its basin, but 
eventually every trajectory ends up in attractor 2. At 
$a\simeq0.115$, another crisis can be observed. In the bifurcation 
diagram (Fig.~\ref{fi.bif5}), the dynamics on the resulting 
attractor resembles that of attractor 1. However, it is not at all 
obvious, whether it is, indeed, attractor 1 or, maybe, the two 
attractors merged. In order to answer this question, we select an 
initial condition on the period-1 orbit of ``attractor 2'' and 
observe how the GRS develops from there. For this purpose, we 
utilize the time-delayed feedback control method introduced by 
Pyragas \cite{Pyr92} (see Appendix, for the utilized numerical 
algorithm). With the aid of the control, we force the GRS onto the 
period-1 orbit of ``attractor 2'' (Fig.~\ref{fi.orbit53}) at 
$a=0.12$ (see Fig.~\ref{fi.kontrolle1}). It should be remarked here 
that, in a strict sense, one can no longer talk of the 
``attractor'' after it has undergone a boundary crisis. However, as 
it still attracts the trajectories of its basin and the transient 
motion along them may be very long, we will, nevertheless, still 
call it attractor. The fact that this is possible (with vanishing 
control signal) demonstrates, that at least the period-1 limit 
cycle of attractor 2 does still exist. Subsequently, we switched 
off the control. First, the trajectory left the unstable periodic 
orbit according to the largest Lyapunov exponent of that orbit. 
Then, it remained on attractor 2 for approximately $1000$ 
revolutions. But, eventually, the trajectory changes very quickly 
to an attractor-1-type shape and, within the patience of the 
authors (for a relatively long time interval of $\Delta t = 
300000$, i.e., about $30000$ revolutions), it never again showed 
attractor-2-type behavior (Fig.~\ref{fi.kontrolle2}). From this, we 
conclude that the attractor at $a=0.12$ is, indeed, attractor~1 and 
attractor 2 has lost its stability. In order to arrive at this 
situation, two things must have happened: attractor~2 must have 
undergone a boundary crisis into the basin of attractor~1 and 
attractor 1 must have separated from the separatrix in an 
inverse boundary crisis. It seems virtually impossible that these 
two incidents occur at the same value of $a$. Thus, depending on 
the order they occur, there should be a small interval of either 
coexisting or merged attractors in between. There is some numerical 
evidence that, indeed, the second situation is valid.

The development of the two attractors as a function of $a$ can also 
be observed in the Lyapunov exponents. Figure~\ref{fi.ly5} shows the
Lyapunov exponents of attractor~1 except for the interval $a\in 
[0.9,0.115]$, where attractor~1 does not exist. The sequence of 
periodicity, quasiperiodicity with periodic windows, and chaos can 
clearly be seen. For $a > 0.115$, the GRS is hyperchaotic 
\cite{Roe79}. In Fig.~\ref{fi.orbit55}, the phase space projections 
of the hyperchaotic attractor at $a=0.3$ are shown. Interestingly, 
the second Lyapunov exponent becomes positive in the vicinity of 
the second and the third crisis. From that point, the three largest 
Lyapunov exponents grow more or less linearly with $a$ (exept for 
a small periodic window at $a\simeq 0.23$). The GRS in this range 
of $a$ can no longer be regarded as a perturbed R"oss\-ler system 
in any projection. Oscillator $(z_1,z_2)$ gradually ceases to be 
dominant and the dynamics of the GRS seems more and more irregular. 
At $a\simeq 0.17$, the third Lyapunov exponent becomes positive. 
The Lyapunov dimension of the attractor is larger than $4$. 
However, no qualitative change of the dynamics could be observed at 
this point, either in the bifurcation diagrams or in the phase space 
projections.

\subsection{Numerical study of the case $N=7$}
We now proceed to show that the GRS does behave in a similar manner 
in the case $N=7$. In the mode picture, the seven-dimensional GRS 
consists of three oscillators coupled via the nonlinear trigger 
$x_7$.

If one compares the bifurcation diagram (Fig.~\ref{fi.bif7}) and the 
corresponding Lyapunov spectrum (Fig.~\ref{fi.ly7}) with those of 
$N=5$ (Figs. \ref{fi.bif5} and \ref{fi.ly5}, respectively), one 
perceives a striking resemblance. There are three coexisting 
attractors for lower $a$ ($a\in[0,0.05]$), corresponding to the 
three oscillators of the seven-dimensional GRS. Two of these 
attractors vanish via boundary crises in an intermediate regime of 
interacting attractors ($a\in[0.06,0.115]$). Finally, only one 
attractor remains, which is hyperchaotic with an increasing number 
of positive Lyapunov exponents.

For $a\in[0,0.05]$, each of the three attractors can be 
characterized by the dominance of one of the three oscillators. 
Each attractor develops as a result of the instability of the 
fixed point $(\vec x^{(1)}, x_N^{(1)})$ corresponding to the 
respective dominant oscillator. This oscillator triggers $x_7$, 
while the other two oscillators have a small amplitude. On 
\textit{attractor 1}, oscillator $(z_1,z_2)$ is dominant, on 
\textit{attractor 2}, oscillator $(z_3,z_4)$, and on 
\textit{attractor 3}, oscillator $(z_5,z_6)$. As long as the 
attractors are periodic (e.g., for $a \in [0.006, 0.04]$ on 
attractor 1) the mode locking is kept up by phase shifts in the two 
oscillators with the small amplitude as in the case $N=5$. Most 
remarkably, on all three attractors, the dynamics is qualitatively 
equal to that of the R"oss\-ler system in the projection onto the 
respective dominant oscillator together with the nonlinear variable 
$x_7$, i.e., in a three-dimensional projection of the 
seven-dimensional phase space (compare Fig.~\ref{fi.orbit7} (a) and
(b) for the period-1 orbit on attractor 2, and Fig.~\ref{fi.orbit7}
(c) and (d) 
for a chaotic orbit on attractor 2). Again, the existence, form, and
location in phase space of the three attractors can be interpreted 
as a manifestation of the structural symmetry. They appear as 
mirror images of each other under the symmetry operation of an 
exchange of the dominant oscillator with one of the other  
oscillators. For example, attractor~1 would be obtained from 
attractor 2 by the transformation $z_1 \leftrightarrow z_3$, $z_2 
\leftrightarrow z_4$. 

For small values of $a$, the dominant oscillator of each of the
attractors has a large amplitude, while the other two oscillators 
remain small. That means, the three attractors are located in phase 
space close to the subspace  of the respective dominant oscillator 
together with the nonlinear trigger $x_7$. Thus, they are clearly 
separate from each other. With increasing $a$, the attractors grow 
and begin to interact. First, attractor~3 vanishes at 
$a\simeq0.052$ in a boundary crisis, where it collides with the 
separatrix to the basin of attractor 1. Attractor 1 also loses 
stability at $a\simeq0.06$, such that only attractor 2 remains 
stable. Finally, at $a=0.11$, a third crisis occurs, which leaves a
large attractor that encompasses the dynamics of all three 
attractors. In general, the motion from one attractor to the other 
is much slower than the motion on the attractors (compare 
Fig.~\ref{fi.kontrolle2}).

In Fig.~\ref{fi.ly7}, the Lyapunov exponents as a function of $a$ 
are shown. For $a < 0.06$, the GRS is on attractor~1. In the 
grey-shaded region $a \in [0.06, 0.115]$, the GRS revolves on 
attractor~2. For $a > 0.115$, the Lyapunov exponents of the 
remaining hyperchaotic attractor can be seen. Here, the Lyapunov 
exponents grow approximately linear with $a$, similar to the case 
$N=5$.

\subsection{Dynamics for arbitrary odd dimension $N$}

We have demonstrated how the dynamics of the GRS in the cases 
$N=5$ and $N=7$ can be understood within the same scheme of 
coexisting attractors. We expect that the dynamics of the GRS 
behaves in an analogous way for every odd dimension $N$. Thus, we 
expect, in general, $N-1\over2$ coexisting attractors for small 
values of $a$.  On each of the attractors, one of the oscillators 
of the mode picture will be dominant. Each attractor will be 
located close to the hyperplane of the respective dominant 
oscillator together with the nonlinear trigger. The other 
oscillators will have a negligible amplitude. In the projection 
onto the dominant oscillator together with the nonlinear trigger 
$x_N$, the dynamics will be R"ossler-like. For higher values of 
$a$, the attractors will grow in phase space and interact with each 
other in several crises. Eventually, one hyperchaotic attractor 
will remain, which extends over all of the phase space that was 
occupied by the $N-1\over2$ coexisting attractors. 

For clarity, we shortly discuss the dynamics of the GRS in phase 
space for a higher number of variables, $N=61$. To this end, we use 
a pseudo space-time representation of the dynamics in 
Fig.~\ref{fi.tra61} for three different values of $a$. The values 
of every second variable $x_{2n+1}, n=0,\ldots,30,$ are depicted in 
grey scale as a function of time. In order to ensure that the 
dynamics has settled on the attractor, the GRS was propagated for a 
transient time 500000, which corresponds to about 10000 revolutions 
of the slowest oscillator for $N=61$. A periodic orbit for $a=0.04$ 
is visualized in Fig.~\ref{fi.tra61}(a). In the mode picture, the 
mean amplitudes of the five dominant oscillators are: $1.30$, $ 
1.16$, $ 1.12$, $ 0.40$, $ 0.11$, $ 0.05$, $ 0.04$. The amplitudes 
of the other oscillators are smaller than $0.01$. Obviously, the 
dynamics is not dominated by a single oscillator, suggesting that 
the coexisting periodic orbits have undergone one or more merging 
crises before losing their stability. In Fig.~\ref{fi.tra61} (b), a 
comparatively low-dimensional chaotic orbit close to the above 
periodic one (Fig.~\ref{fi.tra61}(b)) is shown. It has been 
calculated for $a=0.07$. The trace of the periodic orbit can still 
be seen in the image. Additionally, a considerable activity of 
high-frequency modes can be perceived as well. The largest Lyapunov 
exponents are: $0.00021$, $0.00017$, $0.00014$, $0.00010$, 
$0.00009$, $0.00008$, $0.00008$, $0.00002$, $0.00001$, $-0.00001$, 
$-0.00003$, $-0.00005$. From this, the Lyapunov dimension is 
estimated to be $D=20.8$ with the help of the Kaplan-Yorke 
conjecture \cite{KY79}. In this state, the GRS does not explore all 
of its phase space and several other coexisting chaotic attractors 
can be found, which are not shown here. Figure~\ref{fi.tra61}(c) 
shows a snapshot for $a=0.15$, which lies in the hyperchaotic 
regime. Here, no coherent structure can be perceived. This state 
has 38 positive Lyapunov exponents and a Lyapunov dimension of 
$D=60$. In this state, the hyperchaotic attractor extends to all of 
the phase space and no coexisting attractor can be found.

For large values of $N$, the GRS is similar to a time delay system 
of the form $\dot y(t) = f(y(t-\tau)) + g(y(t))$, as investigated 
,e.g., in \cite{Far82,GMPA94,IM86,IM87,GLP95,BMKP97a,BMKP97b}. This 
can clearly be seen in Fig.~\ref{fi.tra61}. The linear subsystem 
$\vec x$ acts as a delay-line, that merely transports signals back 
and forth between the positive feedback process, represented by 
$x_1$, and the nonlinear trigger $x_N$. The delay time of the GRS 
is approximately $N$ (see~\ref{solutionlinsub}). The difference 
between the GRS and a time delay system is that the GRS -- as a 
spatially discrete system -- has got a nonlinear dispersion 
relation $\omega (k)$. As every mode of the GRS is active for 
values of $a$ in the hyperchaotic regime, the dispersion relation 
cannot be neglected in that case. Thus, the function $f$ would have 
to be replaced by a functional in $y(t-\Delta t)$ with $\Delta t > 
N$. Still, the GRS has got the same basic structure as a time delay 
system. We arrive at identical scaling properties of the Lyapunov 
exponents, the Lyapunov dimension, and the metric entropy for $N 
\rightarrow \infty$, as will be shown in the following section.

\section{Properties of the Generalized R"oss\-ler System under 
         variation of system size $N$}
\label{chap4}
Up to now, we have investigated the dynamics of the GRS in phase 
space under variation of the control parameter $a$ for the cases 
$N=5$ and $N=7$. In the present section, we investigate the GRS as 
a function of $N$ for four fixed values of $a$: $a=0.3, a=0.25, 
a=0.15$, and $a=0.07$. The first three values of $a$ are in the 
regime of large, phase-space filling hyperchaotic attractors, the 
fourth value of $a$ lies in the regime of low-dimensional, 
coexisting attractors. $N$ is varied from $3$ to $61$. Here, we 
mainly focus on the numerical calculation of Lyapunov exponents, 
from which the Lyapunov dimension $D_\lambda$ and the metric 
entropy $H$ are estimated. The Lyapunov exponents have been 
calculated with the fixed initial condition $(\vec x, x_N) = (1, 0, 
\ldots, 0, 0)$. The other parameters are again chosen to be 
$\varepsilon=0.1$, $b=4$, and $d=2$. Interpreting the 
dimensionality $N$ of the GRS as system size, we compare the results 
to the 'thermodynamic limes' of homogeneous, spatially extended 
systems, where the number of positive Lyapunov exponents, the 
Lyapunov dimension, and the metric entropy are reported to be 
proportional to $N$ ~\cite{PPP84,LPR86,EG94,BBW94,CH94}. In this 
case, the distribution of Lyapunov exponents has been reported to 
approach a limit function $f$, i.e., $\lambda_i = f(i/V)$, if $V$ 
is the system size \cite{TPPA91}. In the next step, we draw the 
attention to the GRS's similarity to time-delay systems, where the 
Lyapunov dimension has been reported to be proportional to the 
delay time, while the metric entropy approaches a limit value for 
increasing delay time \cite{Far82}.  

Figure~\ref{fi.lyn1} shows the number of positive Lyapunov exponents
$N^+$, the Lyapunov dimension $D_\lambda$ (estimated via the 
Kaplan-Yorke relation \cite{KY79}) and the metric entropy $H$ 
(estimated via the Pesin formula \cite{Pes77}) as a function of $N$ 
for the four values of $a$. The number of positive Lyapunov 
exponents $N^+$ and the Lyapunov dimension $D_\lambda$ grow 
linearly with $N$. Remarkably, the Lyapunov dimension $D_\lambda$
is maximal up to its fractional part (i.e., $D_\lambda > N-1$) for 
almost all odd dimensions $N$, as long as the GRS is in the 
hyperchaotic regime (large enough $a$). The metric entropy 
approaches a limit value $H_\infty$ for large $N$. Therefore, the 
Lyapunov exponents decrease proportional to $1/N$. Note that this 
observation is not so clear-cut for $a=0.07$, probably due to the 
fact that the chosen initial condition lies within the basin of 
attraction of different attractors for different values of $N$.
Therefore, we find that the limes $N \rightarrow \infty$ of the GRS 
is similar to the limes of an increasing delay time of time-delay 
systems and is not similar to the 'thermodynamic limes' of 
homogeneous, spatially extended systems. Recently, the constancy of 
the metric entropy of time-delay systems for increasing delay time 
has been attributed to the constancy of the number of 'localized 
nonlinearities' \cite{BMKP97b}. The same idea applies to the GRS. 
In Section II, we have distinguished between a linear, 
$(N-1)$-dimensional subsystem $(x_1, ..., x_{(N-1)})$ and a 
one-dimensional nonlinear subsystem (the trigger variable $x_N$).
The linear subsystem $(x_2, ..., x_{(N-1)})$ (without the positive 
feedback process present in $x_1$) allows only for a linear, 
bi-directional transport of signals and is, therefore,
not able to initiate any unstable behavior (see 
Fig.~\ref{fi.tra61}). In the case of the GRS, it is the 
one-dimensional 'localized nonlinearity' $x_N$ \cite{BMKP97b} 
(together with $x_1$), which allows for a nonlinear stretching and 
folding in phase space and is solely responsible for the positive 
metric entropy. As in the case of time-delay systems, we expect the 
value of the metric entropy $H$ for increasing $N$ to depend only 
on the number of localized nonlinearities, which is equal to one 
for the GRS, and the rate with which the information is processed, 
which is estimated via the correlation time $\tau_c$. Therefore, we 
hypothesize that the metric entropy scales like
\begin{equation}
\label{Hlimit}
	H \propto \frac{1}{\tau_c}, 
\end{equation}
for $N \rightarrow \infty$. Since in the hyperchaotic case the 
correlation time is expected to be independent of $N$ (see 
Sec.~III.A), eq.~(\ref{Hlimit}) reproduces the limit behavior of the 
metric entropy of the GRS. In the case of homogeneous, spatially 
extended systems, we expect the number of localized nonlinearities 
to increase proportional with the system size. This leads to 
a proportional increase of the metric entropy with the system size, 
as has been observed in several models. 

Thus, we conjecture that the high-dimensional chaotic motion 
observed in the GRS is fundamentally different compared to the 
spatio-temporal chaos observed in homogeneous, spatially extended 
systems. Here, the difference is expressed in terms of the number 
of localized nonlinearities a system possesses in phase space. In 
the case of the GRS, the number of the localized nonlinearities is 
independent of the control parameters including the system size. In 
the case of homogeneous and nonlinear spatially extended systems,
the number of localized nonlinearities is expected to grow 
proportional to the system size.  

In Table~\ref{tabelle}, we show the limit value $H_\infty$ for 
different values of $a$. Additionally, the value of $H_\infty$ 
normalized to the autocatalytic coefficient $a$ is shown. Note that 
the value of $a$ is an upper bound for the sum of the positive 
Lyapunov exponents and, with this, to the estimated metric entropy 
$H$. The case $a=H_{\infty}$ corresponds to an unbounded, linear, 
$(N-1)$-dimensional system, $(x_1,x_2,...,x_{N-1})$ in absence
of the localized nonlinearity $x_N$. 

From Table~1, we infer that the limit value of the metric entropy, 
$H_\infty$, comes closer to its upper bound $a$ for an increasing 
value of $a$, indicating that the hyperchaotic attractors 
increasingly exploit all the unstable directions available.

In the following, we examine the distribution of Lyapunov exponents 
as a function of $N$. If one is to compare the Lyapunov exponents of
different systems, one faces the problem that the absolute values 
of them depend on the chosen time scale. In systems with many 
different characteristic times, like the GRS, it is not obvious 
which time scale to use. Above, the time scale for the GRS under 
variation of $N$ has been chosen such that the highest frequency of 
the linear subsystem $\omega_{max}$ is approximately equal to $2.0$ 
(see eq.~(\ref{frequenzen})). In this case, we expect the 
correlation time $\tau_c$ to be approximately constant for 
different values of $N$, while the time $\tau_s$ a signal needs to 
traverse the linear subsystem $(x_1, ..., x_{(N-1)})$ increases 
proportional to $N$. In the analysis to follow, we choose the time 
scale such that the rescaled mean positive Lyapunov exponent is 
equal to one. The rescaled Lyapunov exponents $\lambda_s$ are 
defined as
\begin{equation}
    \label{rescaledlambda}
    \lambda_s = {N^+ \over \sum \lambda^+} \lambda\textrm{,}
\end{equation}
where $\sum \lambda^+$ denotes the sum over the positive Lyapunov 
exponents, i.e., the metric entropy, and $N^+$ denotes the number 
of positive Lyapunov exponents. Using the time scale according to 
eq.~(\ref{rescaledlambda}), the maximal frequency of the GRS scales
like $\omega_{max} \approx 2 \frac{N^+}{H}$. Assuming $N^+ \approx 
N$ and $H \rightarrow H_{\infty}$ for high enough $a$, we find
$\omega_{max} \propto \frac{N}{H\infty}$. From this the correlation 
time is estimated to decrease with $N$ in the hyperchaotic case like
$\tau_{c} \propto \frac{H\infty}{N}$. The same argument applies to 
the scaling of the velocity of signals yielding $v \propto 
\frac{N}{H\infty}$. This leads us to observe that the rescaling of 
the time according to eq.~(\ref{rescaledlambda}) makes the time 
$\tau_{s}=N / v$ a signal needs to traverse the linear subsystem
independent of the system size $N$ for sufficiently large values of 
$a$.

In Fig.~\ref{fi.lyn2}, the distribution of the rescaled Lyapunov 
exponents is shown for different values of $N$ for $a=0.25$. The 
Lyapunov exponents are sorted in descending order. The distributions 
converge to a limit distribution, $\lambda_{s,i} = f(i/N)$, with 
increasing $N$. We observe qualitatively the same behavior for all 
values of $a$. Note that, although the existence of a limit 
distribution $f$ has also been shown for homogeneous, spatially 
extended systems, the limes of an increasing system size $N$ for 
the rescaled Lyapunov exponents (\ref{rescaledlambda}) is 
considerably different from the thermodynamic limes of spatially 
extended systems. In the latter case, the correlation time is 
approximately independent of the system size, while in the case of 
the GRS the correlation time decreases proportional to the system 
size $N$.

In delayed dynamical systems with an expansive local dynamics 
(local in time), there has been reported the existence of 
``anomalous'' Lyapunov exponents \cite{LGPA93,GLP95} which do not 
scale like $1/N$ for $N\rightarrow \infty$, but remain at a finite 
value. Considering the similarity of the GRS with a delayed system 
and the expansive term present in $\dot x_1=a x_1 - x_2$, one may 
expect to observe such an anomalous Lyapunov exponent for $a>0$. 
However, due to the fact that the linear subsystem transports 
energy (in the form of squared amplitudes) away from $x_1$, the 
local dynamics of the beginning of the linear subsystem is that of 
a damped oscillator for all values of $a$ considered in this paper. 
For values of $a$ being sufficiently large such that the local 
dynamics does indeed become expansive, the GRS is globally 
unstable, as the nonlinear trigger is no longer able to keep the 
amplitudes of the linear subsytem bounded. Accordingly, no sign of 
an anomalous Lyapunov exponent (i.e., a Lyapunov exponent that does 
not scale like $1/N$) could be seen in any of the spectra 
calculated for the present work.

Finally, we would like to discuss the shape of the limit 
distribution $f$ of the rescaled Lyapunov exponents. We have 
observed above that the limit value of the metric entropy, 
$H_{\infty}$, approaches its upper bound $a$ for an increasing 
value of $a$ as a direct consequence of the dynamics increasingly
exploiting all unstable directions of the unstable fixed point 
$(\vec x^{(1)}, x_N^{(1)})$. We find that the same idea applies to 
the Lyapunov spectra: For an increasingly hyperchaotic attractor 
(increasing value of $a$), the Lyapunov spectra gain similarity with
the real parts of the eigenvalues of the unstable fixed point $(\vec 
x^{(1)}, x_N^{(1)})$. In Fig. 18, we show the Lyapunov spectra for 
three different values of $a$ (fixed dimension $N=61$) together with
the real parts of the eigenvalues of the fixed point according to 
eq.~(\ref{frequenzen}). In all cases, the time scale has been chosen 
such that the trace of the matrix $A$ equals one.

\section{Discussion and Conclusion}
We have investigated the GRS as a model for high-dimensional
chaos as the latter emerges out of low-dimensional chaos. One 
important feature of the GRS is that it consists of a linear 
subsystem with a variable number of degrees of freedom together 
with one nonlinear trigger. The linear subsystem can be solved 
analytically. Utilizing the eigenmodes of the linear subsystem, one 
can transform the GRS into a mode picture, consisting of harmonic 
oscillators that are coupled only via the nonlinear trigger. The 
mode picture reveals a structural symmetry of the GRS. With the aid 
of this structural symmetry, we interpret the dynamics of the 
cases $N=5$ and $N=7$ within a general scheme of coexisting 
attractors. For small values of $a$, there is, for each attractor, 
a specific projection into a three-dimensional subspace, where the 
dynamics of the GRS is R"ossler-like. The attractors expand  with 
increasing $a$ and interact in several crises. This parameter 
regime of interacting attractors eventually leaves one large 
hyperchaotic attractor with many positive Lyapunov exponents. Even 
such kind of hyperchaotic dynamics can be made accessible to the 
human mind, which is used to envisage in three spatial dimensions, 
with the help of three-dimensional projections onto the oscillators 
of the mode picture.

In the second part of this paper, we have investigated the Lyapunov 
exponents and related chaotic indicators of the GRS in the limit of 
large values of $N$, mainly in the hyperchaotic regime. The number 
of positive Lyapunov exponents and the Lyapunov dimension grow 
linearly with $N$. The Lyapunov dimension is maximal, $D_{\lambda} 
\approx N$, independently of $a$, as long as one is in the 
hyperchaotic regime. The metric entropy converges to a limit value 
for increasing $N$. If the time is rescaled in such a way that the 
signal travelling time through the linear subsystem remains constant
with increasing $N$, the distribution of the Lyapunov exponents 
approaches a limit function and the metric entropy grows linearly 
with $N$ for $N \rightarrow\infty$. We have argued that the 
hyperchaotic dynamics observed in the GRS is fundamentally 
different from spatio-temporal chaos. In this paper, we have 
expressed this difference in terms of the number of localized 
nonlinearities. 

In Section I, we have raised the question how dynamical systems 
develop from low-dimensional chaotic behavior to hyperchaotic 
states. In the GRS, one observes one specific path through the 
chaotic hierarchy, starting from a stable fixed point over chaos up 
to hyperchaos. The GRS exhibits a scenario of coexisting 
R"ossler-like attractors that interact and eventually merge
to form a hyperchaotic attractor. We would like to emphasize that 
the GRS only realizes one possible way, ending up with one special 
form of hyperchaos. We feel that there are many different forms of 
hyperchaos which possibly cannot be sufficiently characterized with 
the help of Lyapunov exponents. In the case of the GRS, the 
structure of the system has provided a helpful scheme to 
interpret the dynamics. We believe that the investigation of the  
topological structure of the flow and, specifically, the attractor 
structure and the interaction of attractors, could be used to 
classify high-dimensional chaotic dynamics in general.

\section*{Acknowledgements}
We thankfully acknowledge fruitful discussions with G.~Baier, 
S.~Sahle, J.~Peinke, and O.E.~R"ossler. The valuable contribution
of the referee is worth mentioning.
The present work has been supported 
financially by the Deutsche Forschungsgemeinschaft.

\begin{flushleft}

\end{flushleft}

\appendix
\section*{Numerical algorithms applied}
\label{appendixa}

The differential equations were integrated using a Runge-Kutta 
triple~\cite{DP86}. This algorithm takes advantage of a sixth-order 
formula to propagate a system of ordinary differential equations. A 
fifth-order formula is used to estimate the integration error and, 
additionally, for each Runge-Kutta step $t_n \rightarrow t_{n+1}$, 
the algorithm calculates a polynom that approximates the solution 
on the whole interval $[t_n,t_{n+1}]$ up to an error of fifth 
order. The tolerance was set to $10^{-10}$. 

The Poincar' sections were obtained via parabolic interpolation in 
the vicinity of the intersection points. To calculate the bifurcation
diagrams, we utilized a simple algorithm. For each value of $a$, we 
started with the state of the system for the last value of $a$, let 
the system adjust to the changed parameter value for a transient 
time equivalent to some thousand revolutions, and then recorded the 
intersection points.

Lyapunov exponents were calculated using the algorithm described in 
\cite{SN79,BGGS80,PC89}. This algorithm tracks the time development 
of an orthonormal basis in the tangent space of phase space. At 
regular time intervals ($\Delta T=50$ in our case), the vectors are 
reorthonormalized. The mean logarithmic growth rates of the moduli 
of the vectors are the Lyapunov exponents.

The delay equation of the time-delayed control was integrated by the
same Runge-Kutta triple, using the spline polynoms that the 
Runge-Kutta triple outputs to record the continuous history of 
$(\vec x, x_N$). The control acts on all variables of the system. 
We choose a gain factor of $0.05$ and limited the control signal to 
$25\%$ of the modulus of the corresponding time derivative of the 
uncontrolled GRS. For the delay time $\tau$, we choose the 
eigenfrequency of the oscillator to be stabilized and, 
subsequently, adjusted $\tau$ to minimize the mean control signal.

%
\begin{figure}[!ht]
\caption[]{Schematical representation of the structure of the GRS in
           (a) the Baier-Sahle picture,
           (b) the mode picture.
Each circle represents one linear degree of freedom, the square
represents a nonlinear degree of freedom.}
\label{fi.trafo}
\end{figure}
\begin{figure}[!ht]
\caption{The real parts of the eigenvalues of the central fixed 
point $(\vec x^{(1)}, x_N^{(1)})$ close to the Hopf bifurcations 
that give rise to the coexisting attractors ($N=7$) plotted as a 
function of parameter $a$.}
\label{fi.fista}
\end{figure}
\begin{figure}[!ht]
\caption[]{Bifurcation diagram for $N=5$ under variation of $a$. 
Shown is the projection of the Poincar\'e section onto $z_2$ 
(intersection at $z_1 = 0$, parameters: $\varepsilon=0.1$, $b=4$, 
$d=2$).}
\label{fi.bif5}
\end{figure}
\begin{figure}[!ht]
\caption[]{Lyapunov spectrum for $N=5$ under variation of $a$. The
diagram shows the Lyapunov exponents $\lambda$ on attractor~1, 
except for the grey-shaded area, where attractor~1 does not exist 
(parameters: $\varepsilon=0.1$, $b=4$, $d=2$).}
\label{fi.ly5}
\end{figure}
\begin{figure}[!ht]
\caption[]{Period-1 orbit for $N=5$ on attractor 1; 1:2 mode 
           locking; projections onto
           (a) oscillator $(z_1,z_2)$ and $x_5$,
           (b) oscillator $(z_3,z_4)$ and $x_5$
           (parameters: $a=0.04$, $\varepsilon=0.1$, $b=4$, $d=2$).
          }
\label{fi.orbit51}
\end{figure}
\begin{figure}[!ht]
\caption{Chaotic orbit for $N=5$ on attractor 1. Projections onto
              (a) oscillator $(z_1,z_2)$ and $x_5$, 
              (b) oscillator $(z_3,z_4)$ and $x_5$
         (parameters: $a=0.085$, $\varepsilon=0.1$, $b=4$, $d=2$).
         }
\label{fi.orbit52}
\end{figure}
\begin{figure}[!ht]
\caption{Period-1 orbit for N=5 on attractor 2; 2:1 mode locking.
         Projections onto
              (a) oscillator $(z_1,z_2)$ and $x_5$, 
              (b) oscillator $(z_3,z_4)$ and $x_5$
         (parameters: $a=0.04$, $\varepsilon=0.1$, $b=4$, $d=2$).
         }
\label{fi.orbit53}
\end{figure}
\begin{figure}[!ht]
\caption{Chaotic orbit for $N=5$ on attractor 2; projections onto
              (a) oscillator $(z_1,z_2)$ and $x_5$,
              (b) oscillator $(z_3,z_4)$ and $x_5$ 
         (parameters: $a=0.1$, $\varepsilon=0.1$, $b=4$, $d=2$).
         }
\label{fi.orbit54}
\end{figure}
\begin{figure}[!ht]
\begin{tabular}{c}
\begin{minipage}[b]{\the\hsize}
\end{minipage} \\
\begin{minipage}[b]{\the\hsize}
\end{minipage}
\end{tabular}
\caption{The transient onto the period-1 orbit of attractor 2 
         under the action of the time-delay control:  
          (a) projection onto oscillator $(z_1,z_2)$ and $x_5$, 
          (b) projection onto oscillator $(z_3,z_4)$ and $x_5$,
          (c) time development of the amplitude $A$ of the control
              signal $|\vec x(t)-\vec x(t-\tau)|$ 
        (parameters: $N=5$, $a=0.12$, $\varepsilon=0.1$, $b=4$,
         $d=2$; parameters of the control: $\tau=3.8557$,
         control gain~$\kappa=0.05$, control limit~$=0.25$).}
\label{fi.kontrolle1}
\end{figure}
\begin{figure}[!ht]
\caption[]{Time development of the mean amplitudes 
$r_1=\sqrt{z_1^2+z_2^2}$ and $r_2=\sqrt{z_3^2+z_4^2}$ of the two 
oscillators in the mode picture after the time-delay control has 
been switched off. The trajectory remains in the basin of 
attraction of attractor 1 until $t\simeq9500$. At this point, it 
quickly moves into the basin of attraction of attractor 2 and never 
returns to attractor 1 afterwards (parameters: $N=5$, $a=0.12$, 
$\varepsilon=0.1$, $b=4$, $d=2$).}
\label{fi.kontrolle2}
\end{figure}
\begin{figure}[!ht]
\caption{Hyperchaotic orbit for $N=5$; projections onto
              (a) oscillator $(z_1,z_2)$ and $x_5$,
              (b) oscillator $(z_3,z_4)$ and $x_5$ 
         (parameters: $a=0.3$, $\varepsilon=0.1$, $b=4$, $d=2$).
         }
\label{fi.orbit55}
\end{figure}
\begin{figure}[!ht]
\caption[]{Bifurcation diagram for $N=7$: attractor~1, attractor~2,
           and attractor~3 from bottom to top in different
           shadings. The values of $x_6$ at the maxima of $x_6$ are
           shown (parameters: $\varepsilon=0.1$, $b=4$, $d=2$).}
\label{fi.bif7}
\end{figure}
\begin{figure}[!ht]
\caption{Lyapunov spectrum for $N=7$: depicted is up to $a=0.06$ the 
         development on attractor~1, from $a=0.06$ to $a=0.11$ that
         on attractor~2 (grey-shaded), above $a=0.11$ that on the
         remaining large attractor.}
\label{fi.ly7}
\end{figure}
\begin{figure}[!ht]
\caption{(a)-(c): Periodic orbit for $N=7$, $a=0.035$ on attractor~2
                  with 1:1:1 mode locking: projections onto
             (a) oscillator $(z_1,z_2)$ and $x_7$,
             (b) oscillator $(z_3,z_4)$ and $x_7$,  
             (c) oscillator $(z_5,z_6)$ and $x_7$. 
         (d)-(f): Chaotic orbit for $N=7$, $a=0.095$ on attractor~2;
                  projections onto
              (d) oscillator $(z_1,z_2)$ and $x_7$,
              (e) oscillator $(z_3,z_4)$ and $x_7$, 
              (f) oscillator $(z_5,z_6)$ and $x_7$
         (other parameters: $\varepsilon=0.1$, $b=4$, $d=2$).}
\label{fi.orbit7}
\end{figure}
\begin{figure}[!ht]
\caption{Time development of the GRS for $N=61$. The amplitudes of 
the variables $x_{2n-1}$ are shown in grey scale as a function of 
time for (a) $a=0.03$, (b) $a=0.07$, (c) $a=0.15$ (other 
parameters: $\varepsilon=0.1$, $b=4$, $d=2$).}
\label{fi.tra61}
\end{figure}
\begin{figure}[!ht]
\caption{(a) Number of positive Lyapunov exponents, (b) Lyapunov 
dimension, and (c) metric entropy as a function of $N$ for $a=0.3$, 
$a=0.25$, $a=0.15$, $a=0.07$ (other parameters: $\varepsilon=0.1$, 
$b=4$, $d=2$)).}
\label{fi.lyn1}
\end{figure}
\begin{figure}[!ht]
\caption{Distribution of rescaled Lyapunov exponents $\lambda_s = 
{N^+ \over \sum \lambda^+} \lambda$ for different $N$ ($N=15, 21, 
41, 61$). The Lyapunov exponents are sorted in descending order 
(parameters: $a=0.3$, $\varepsilon=0.1$, $b=4$, $d=2$)).}
\label{fi.lyn2}
\end{figure} 
\begin{figure}[!ht]
\caption{Comparison of the real parts of the eigenvalues of the 
linear subsystem $\alpha_i$ and the Lyapunov exponents of the GRS 
in the case $N=61$ for $a=0.07$, $a=0.15$, and $a=0.25$. Here, all 
values have been divided by the respective values of $a$, in order to
get comparable values for different $a$. (other parameters: 
$\varepsilon=0.1$, $b=4$, $d=2$).}
\label{fi.lyn3}
\end{figure}

\begin{table}
\begin{tabular}{@{\hspace{.4in}}ddd@{\hspace{.4in}}} 
a & $H_\infty$ & $H_\infty \over a$ \\ \hline
0.3  & 0.2  & 0.7 \\ 
0.25 & 0.16 & 0.6 \\ 
0.15 & 0.08 & 0.5 \\ 
0.07 & 0.02 & 0.3 \\ 
0.03 & 0.0  & 0.0 
\end{tabular}
\caption{Limit value of the metric entropy $H$ for different values 
of $a$.}
\label{tabelle}
\end{table}

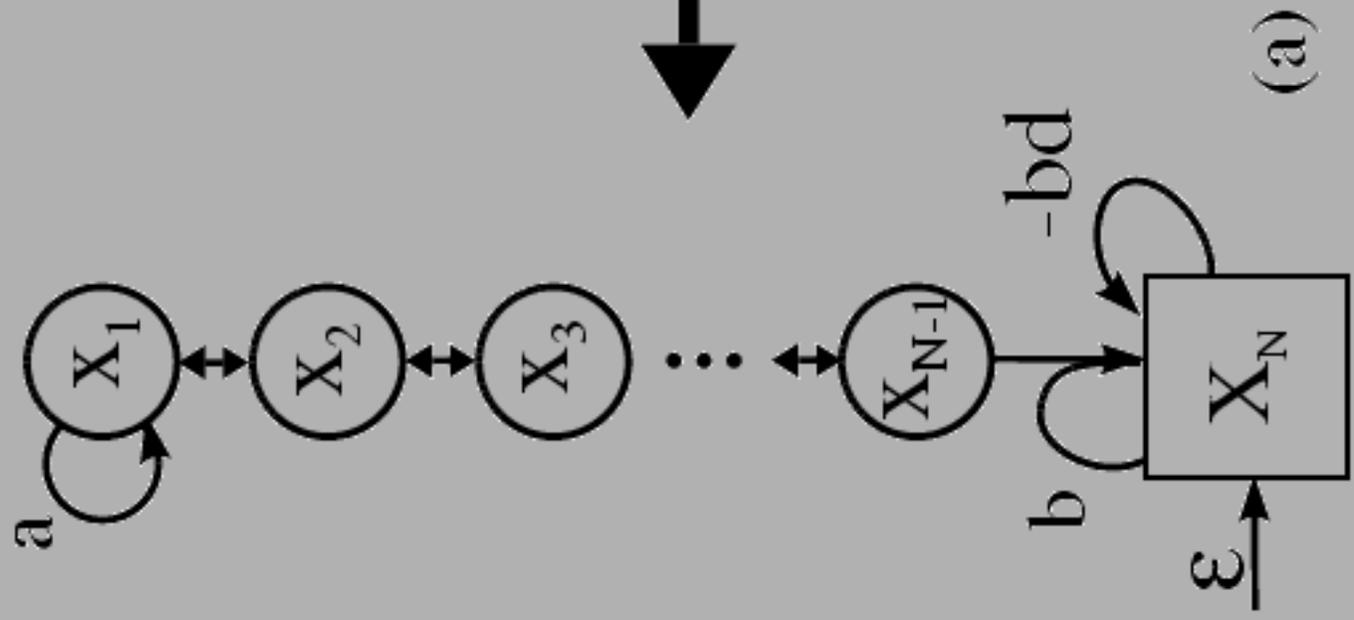
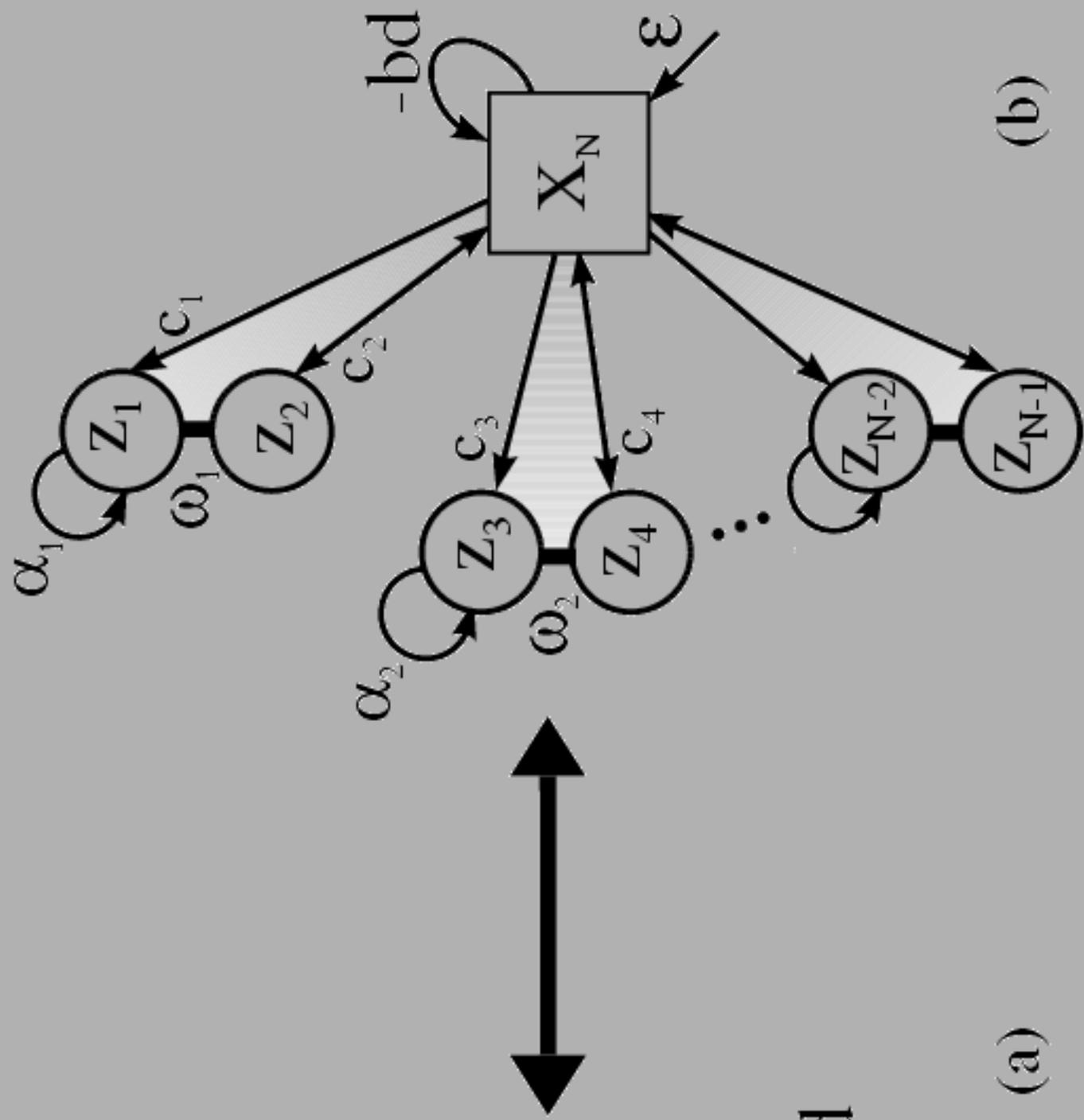
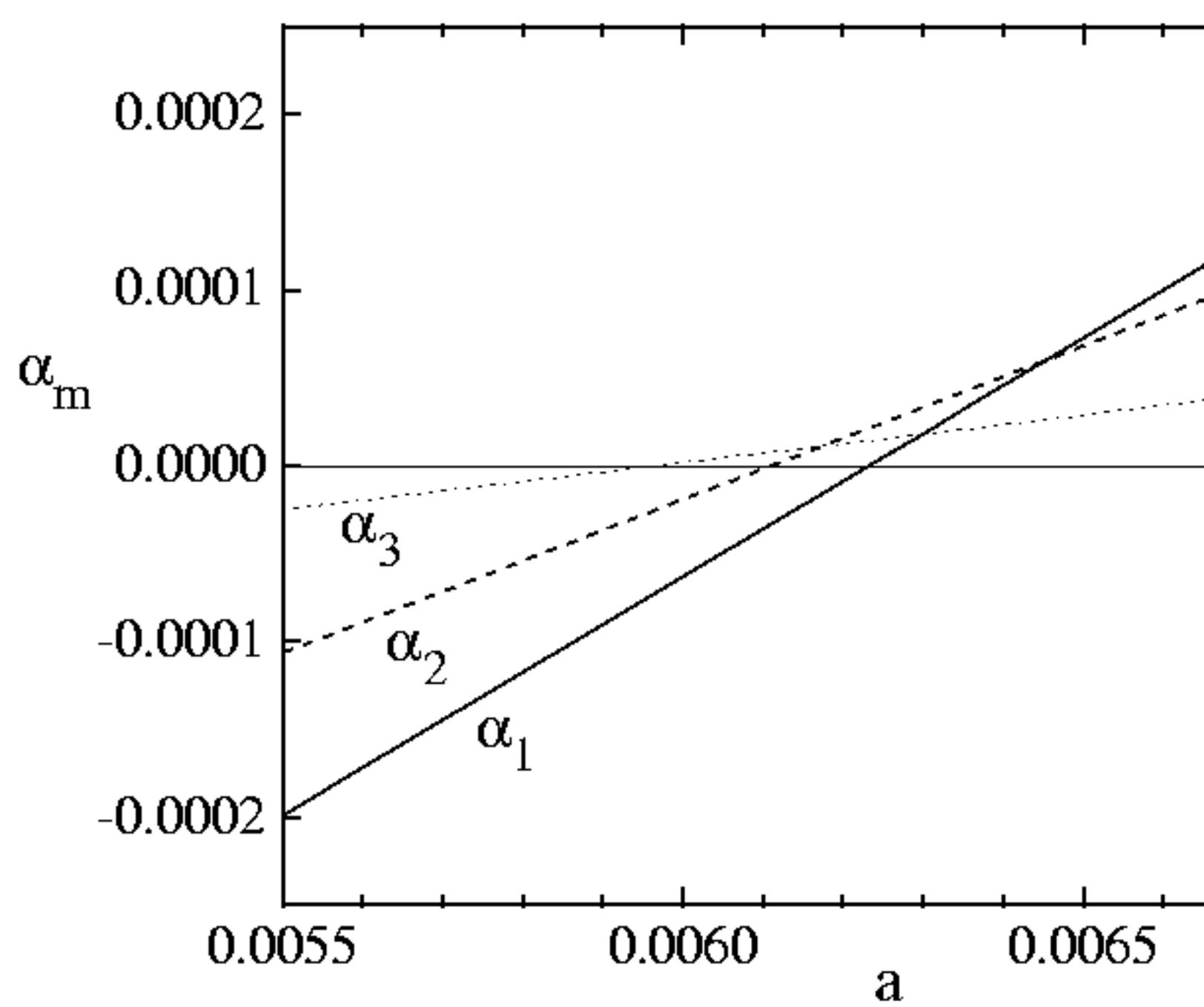
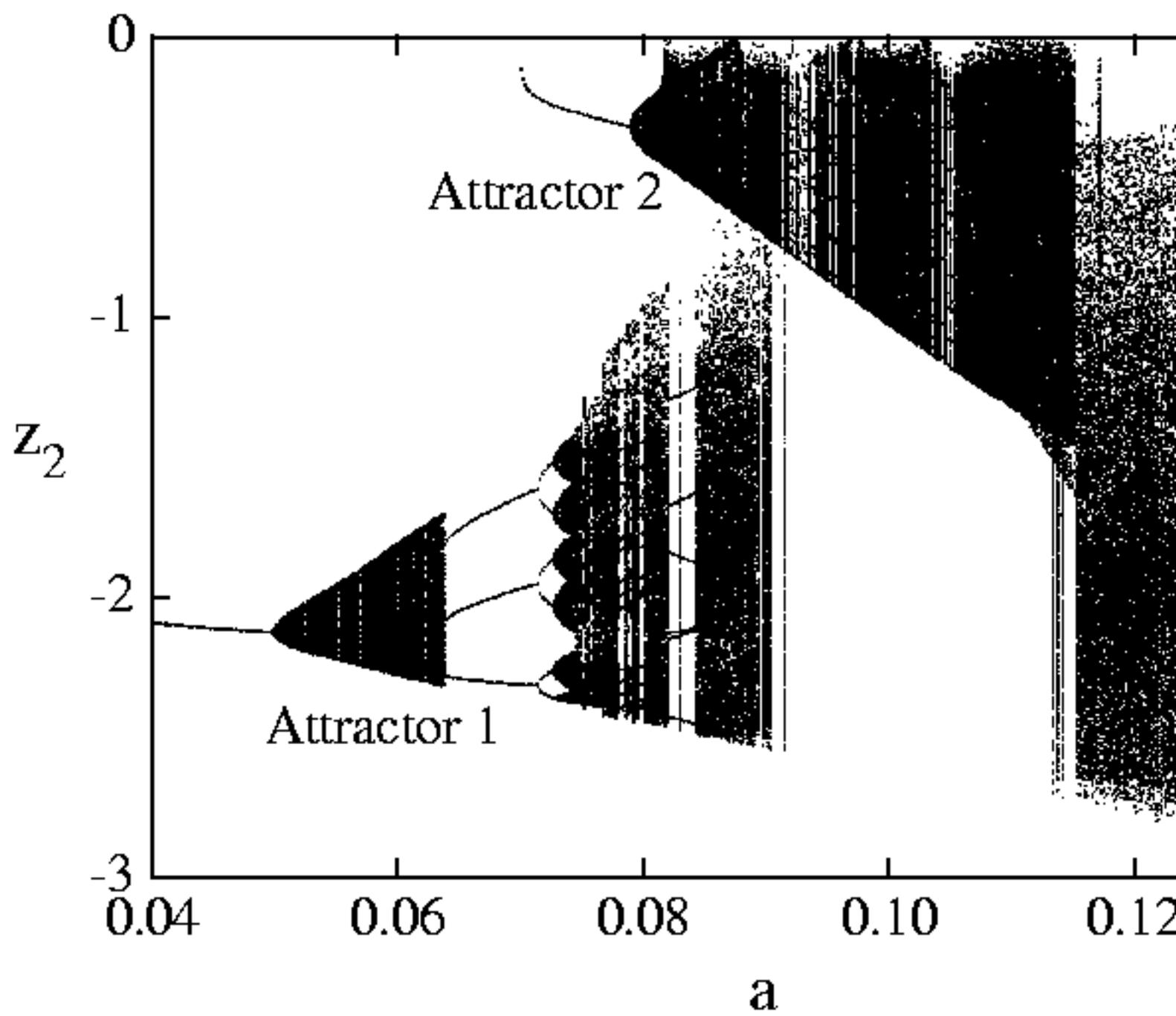
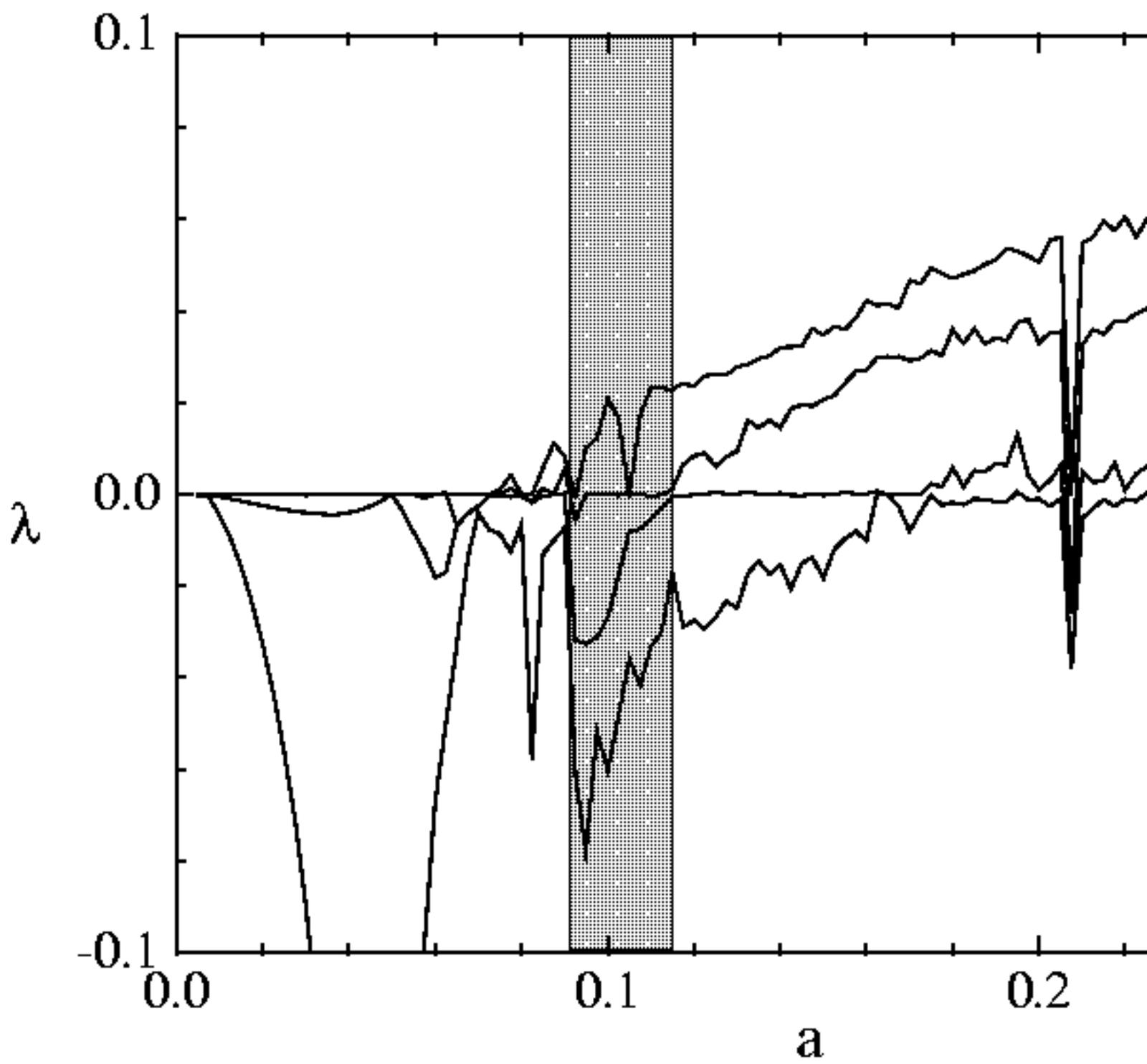
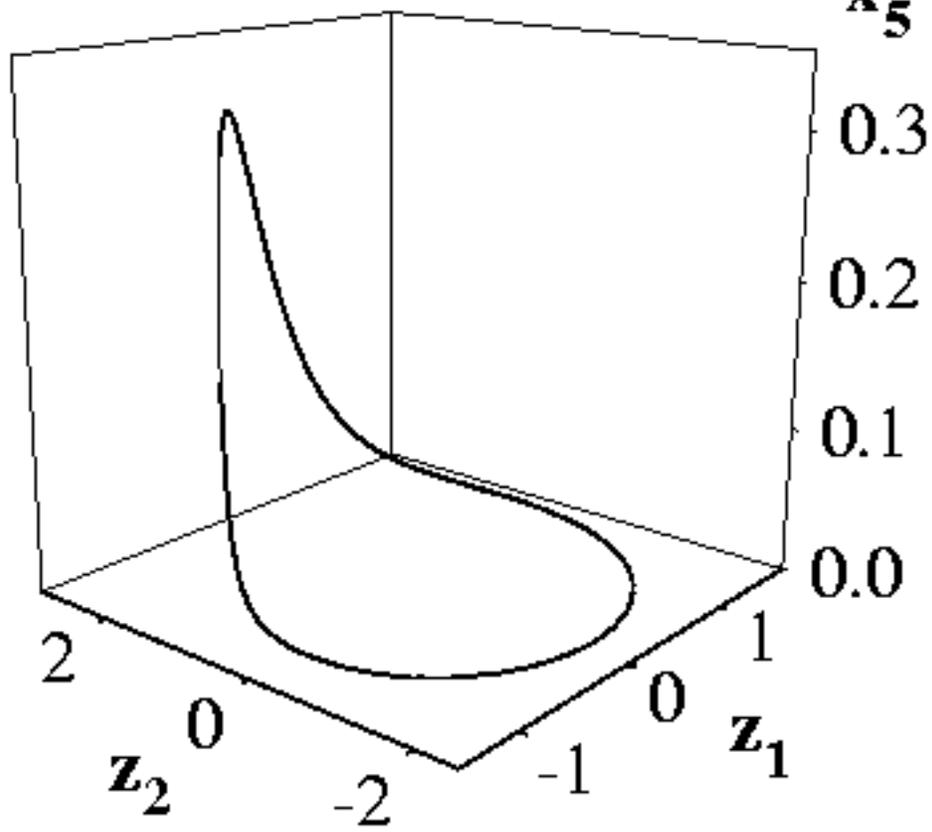
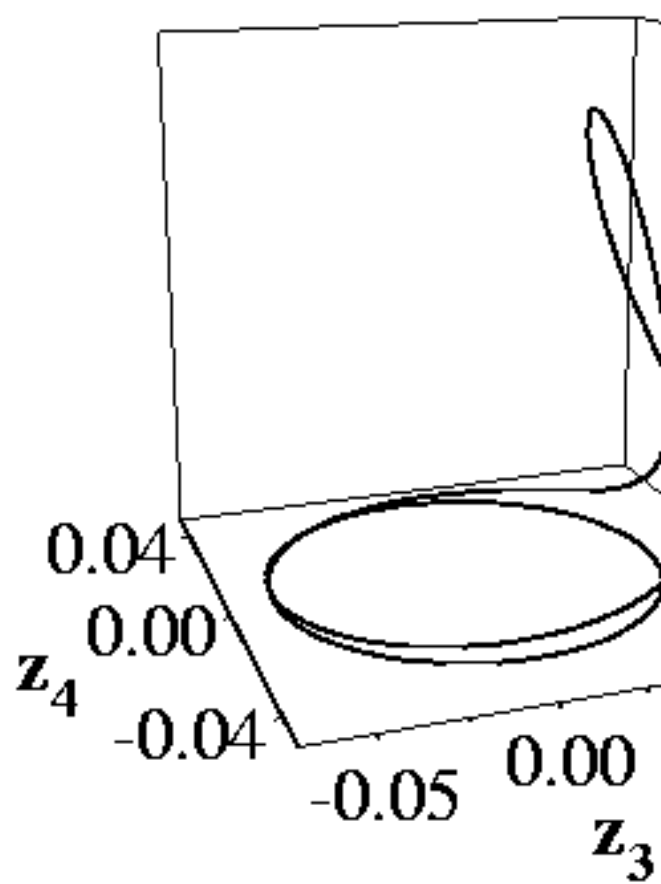
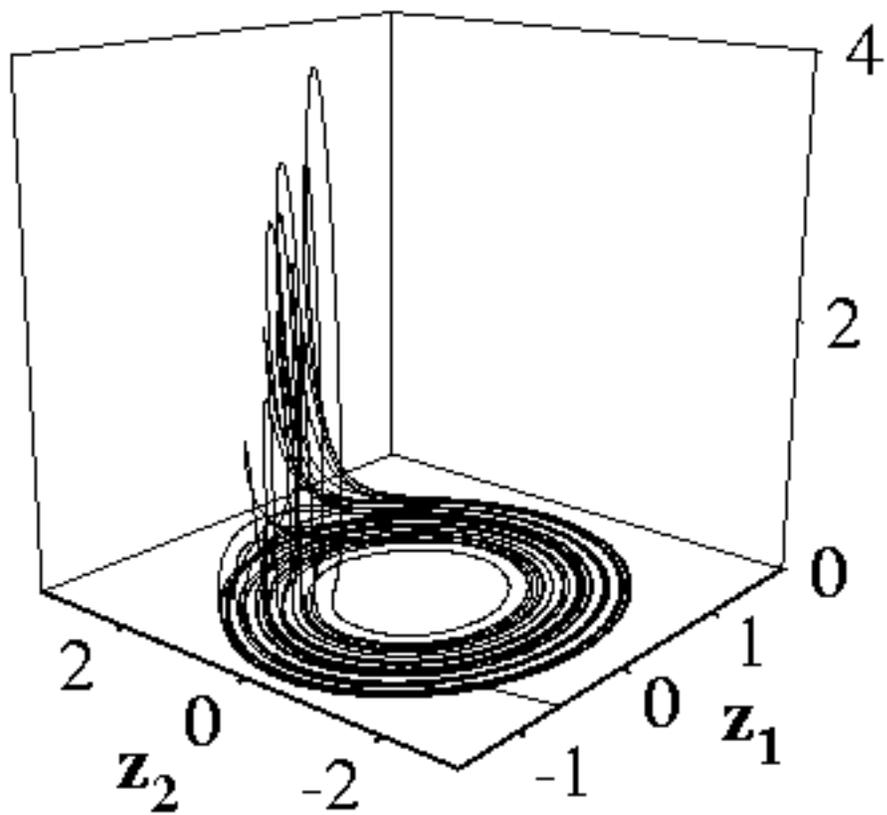
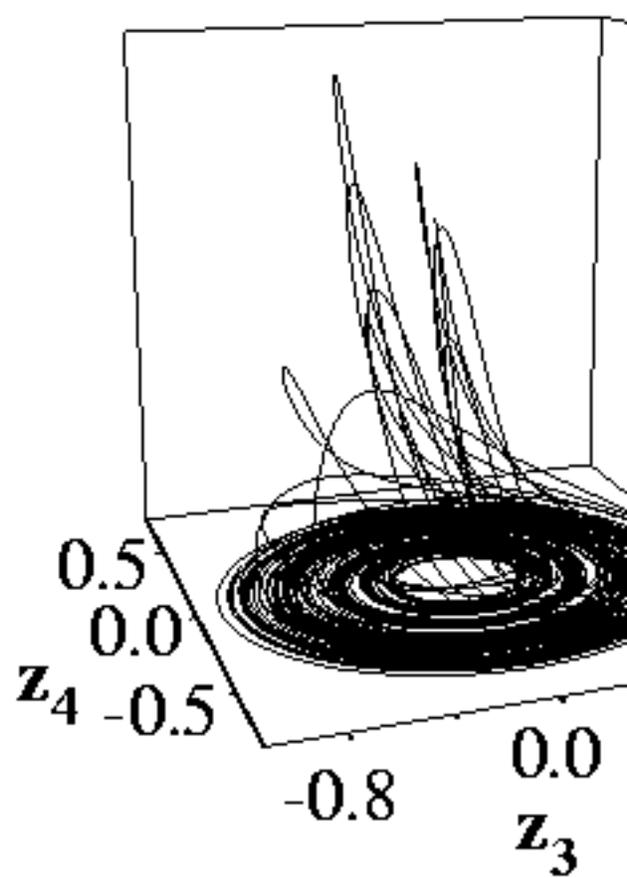
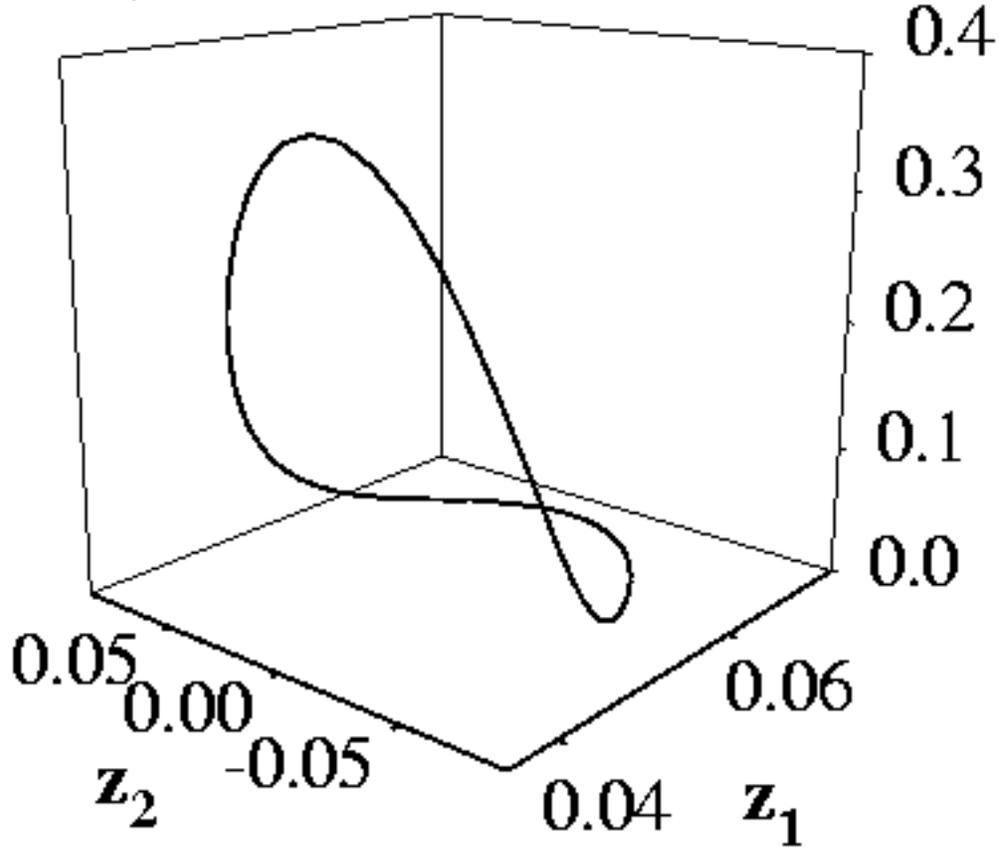
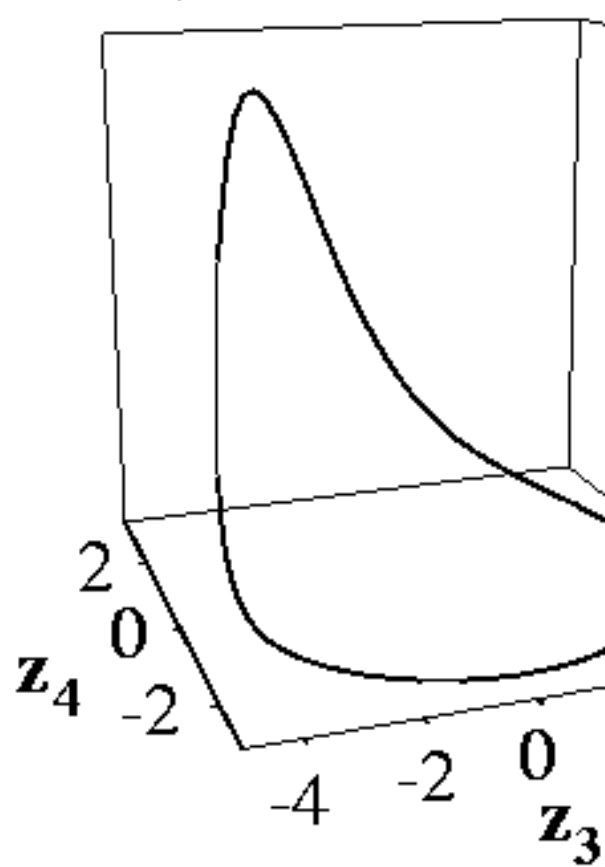
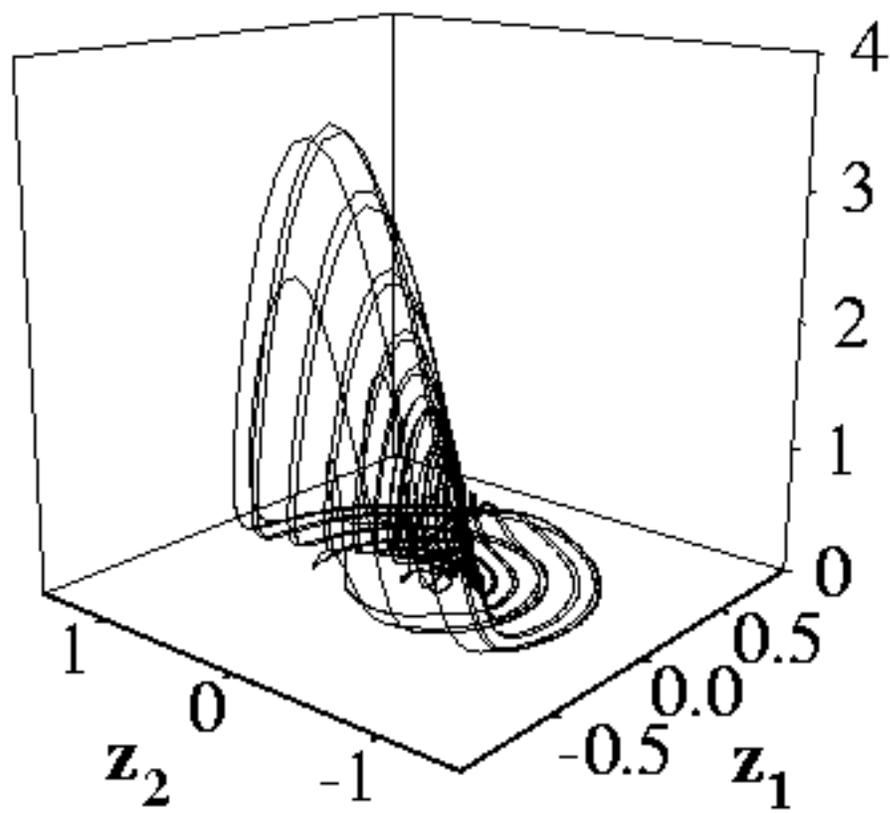
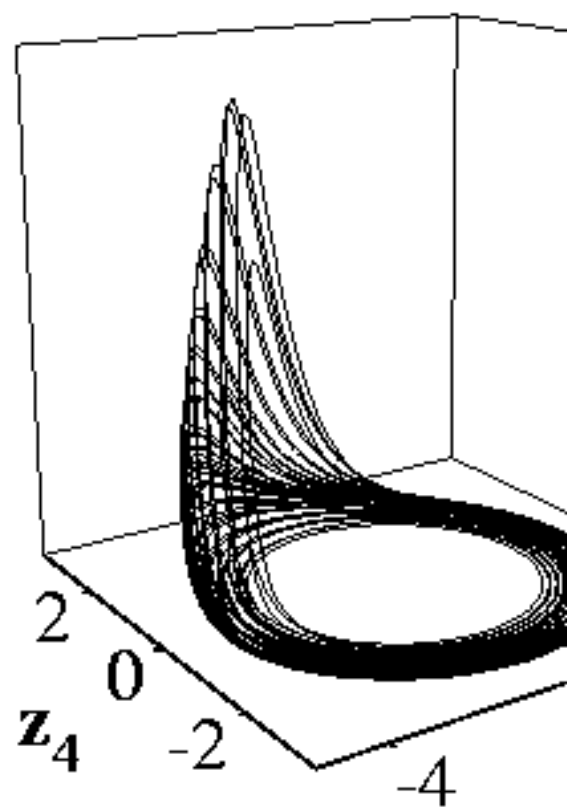
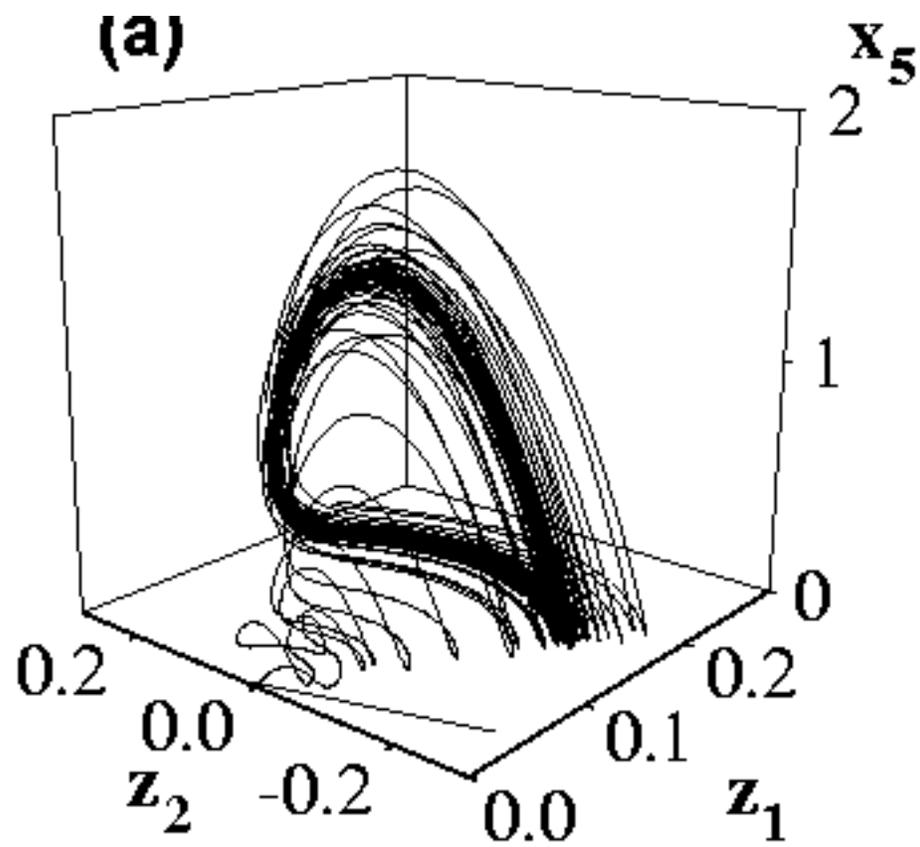
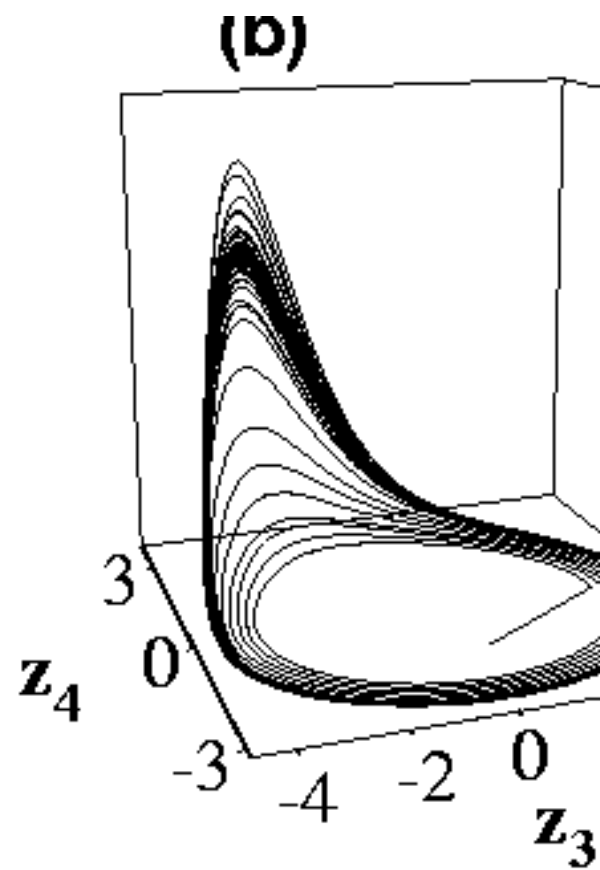
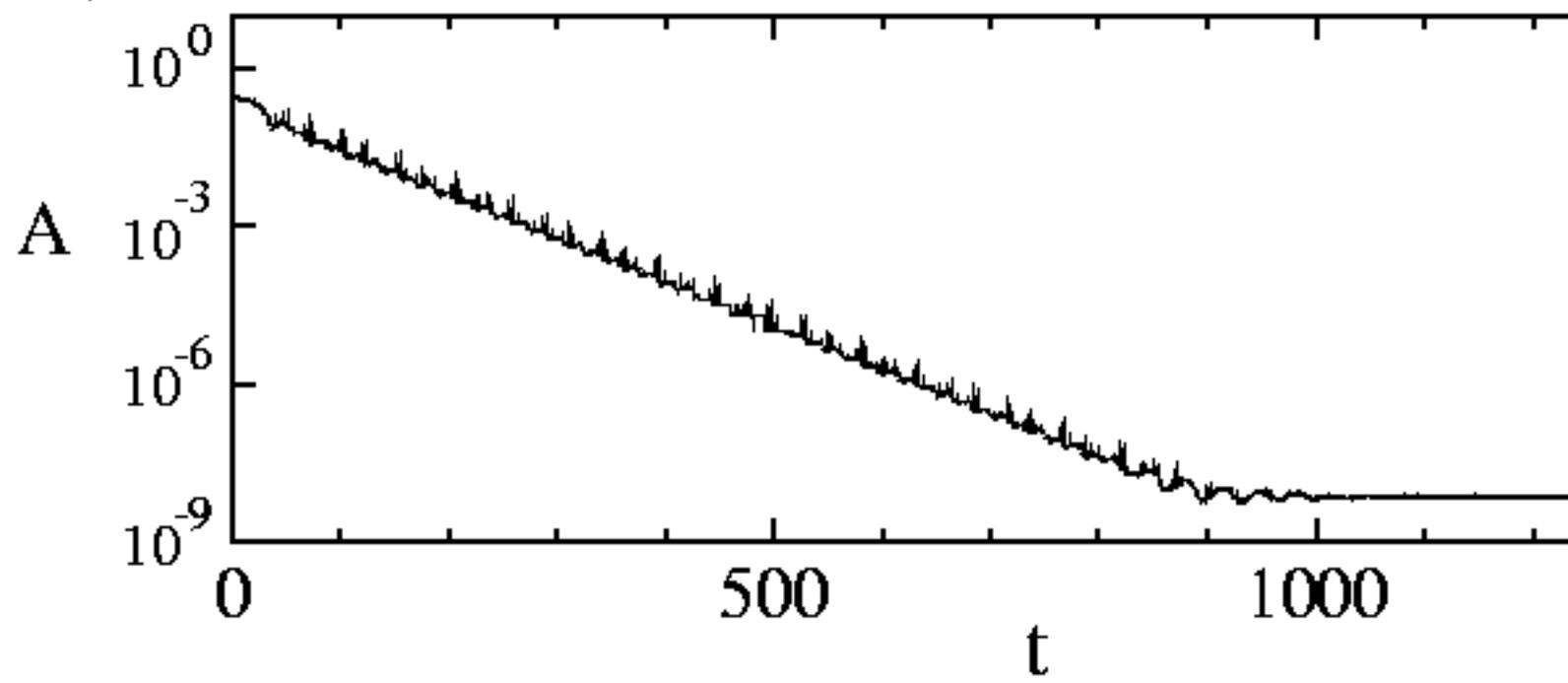
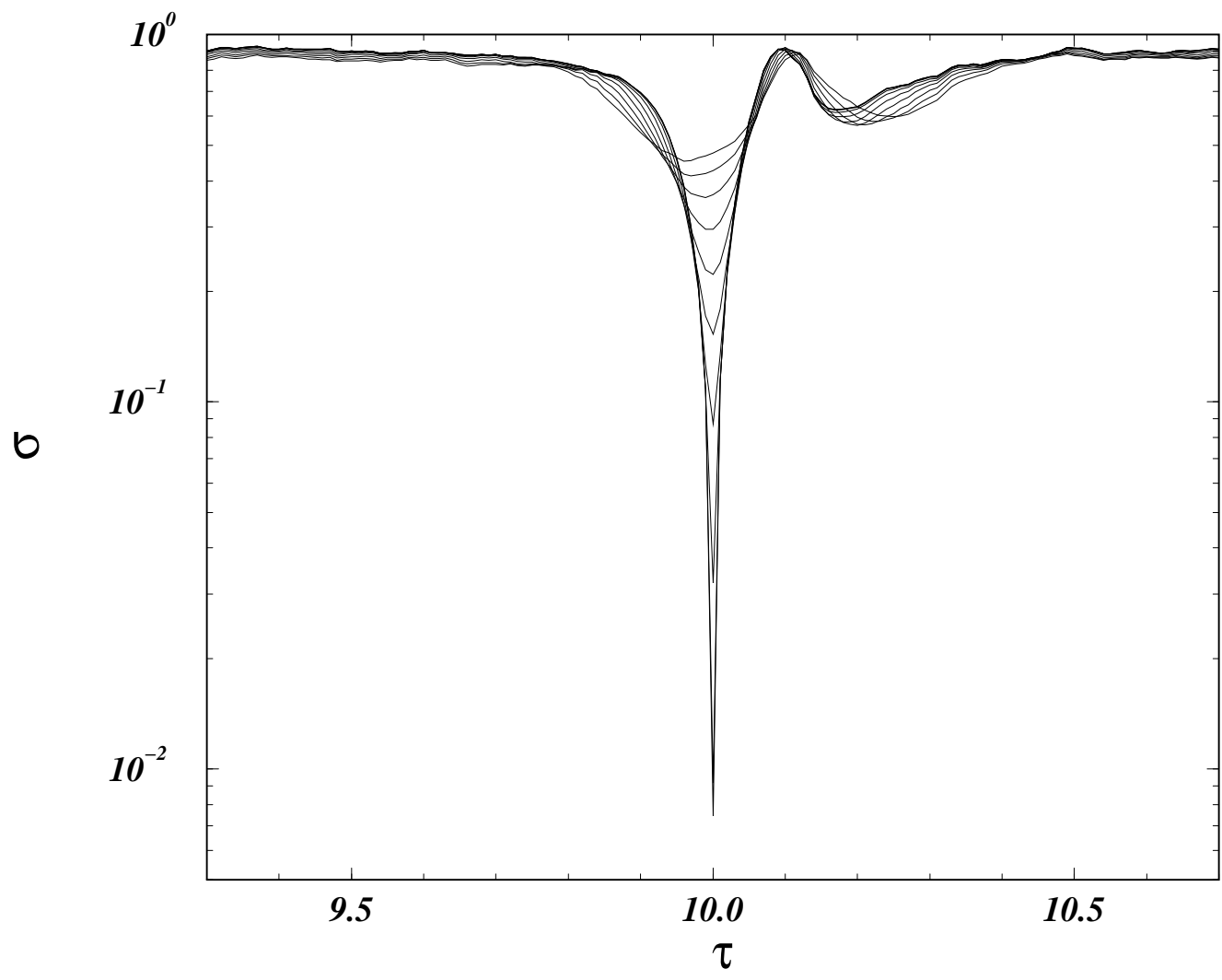
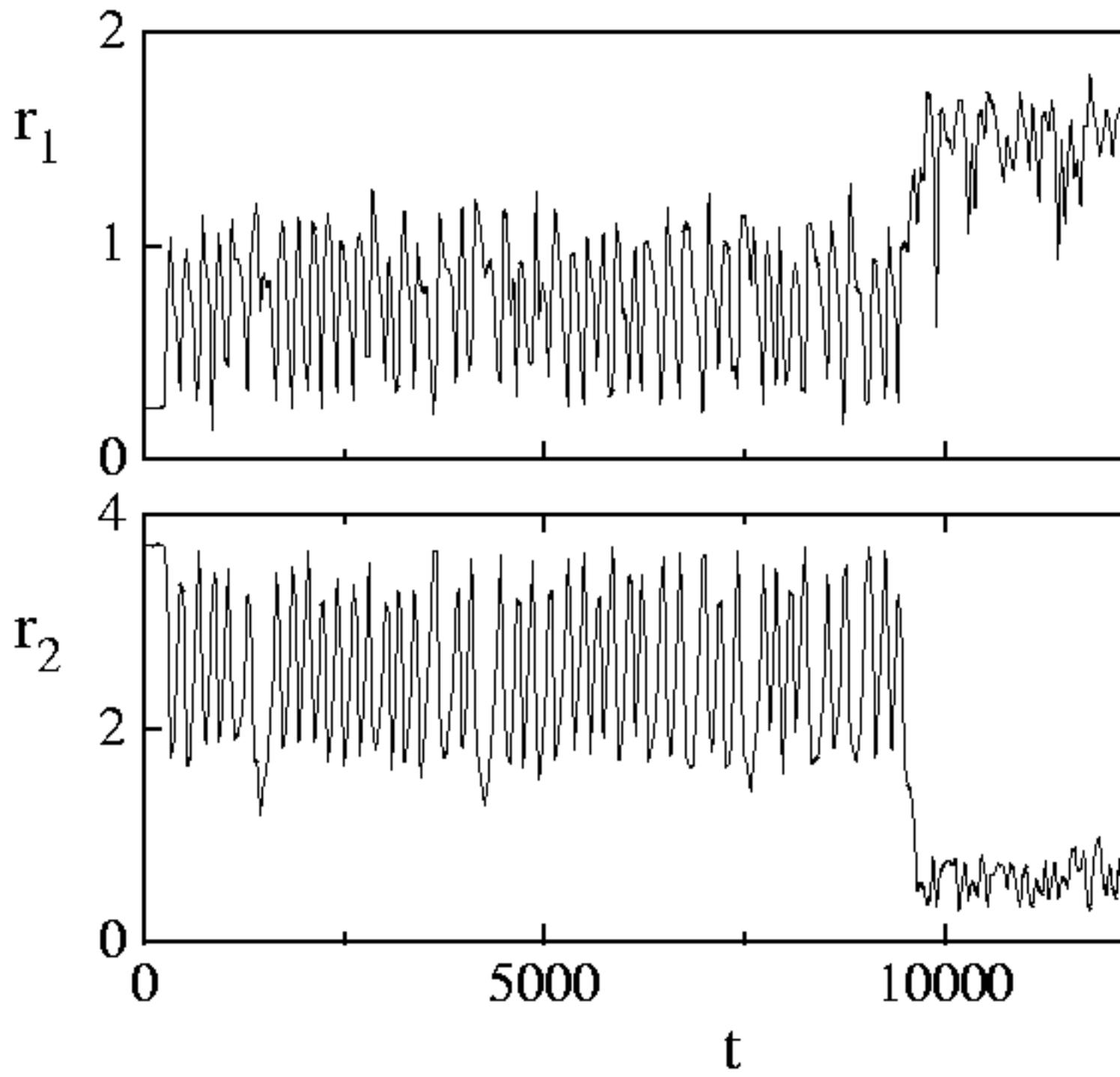
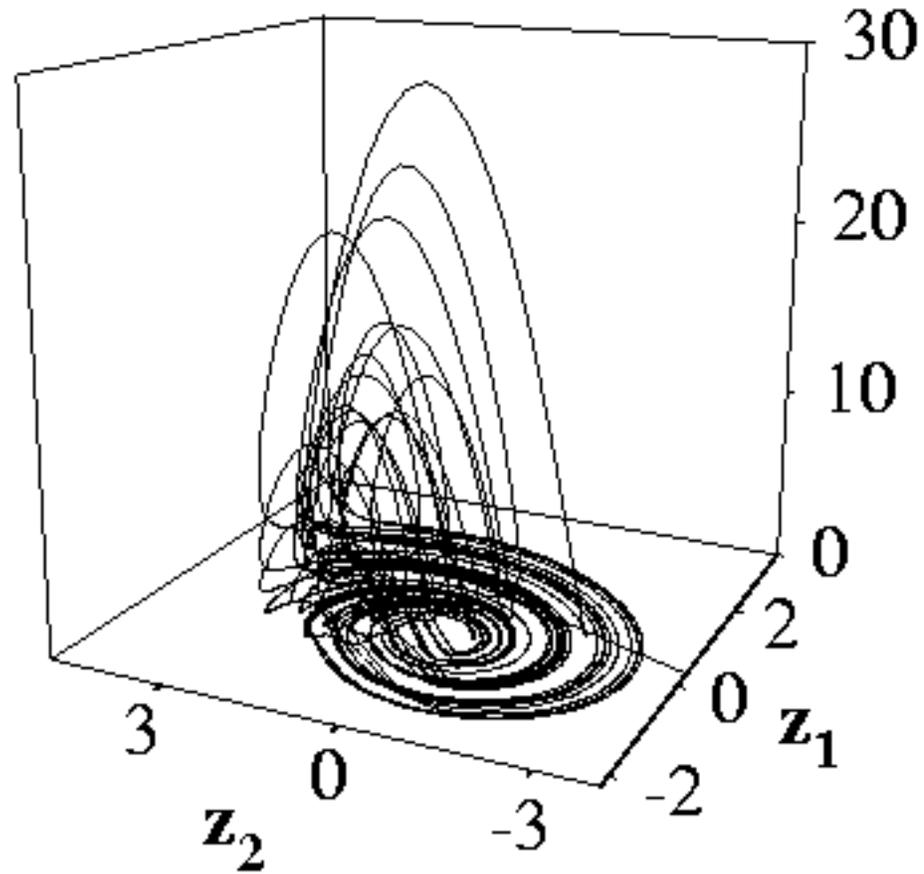
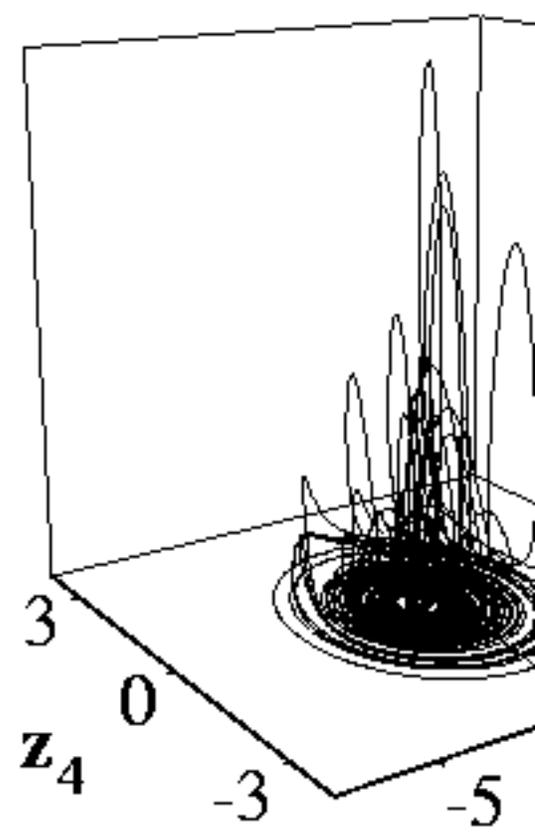
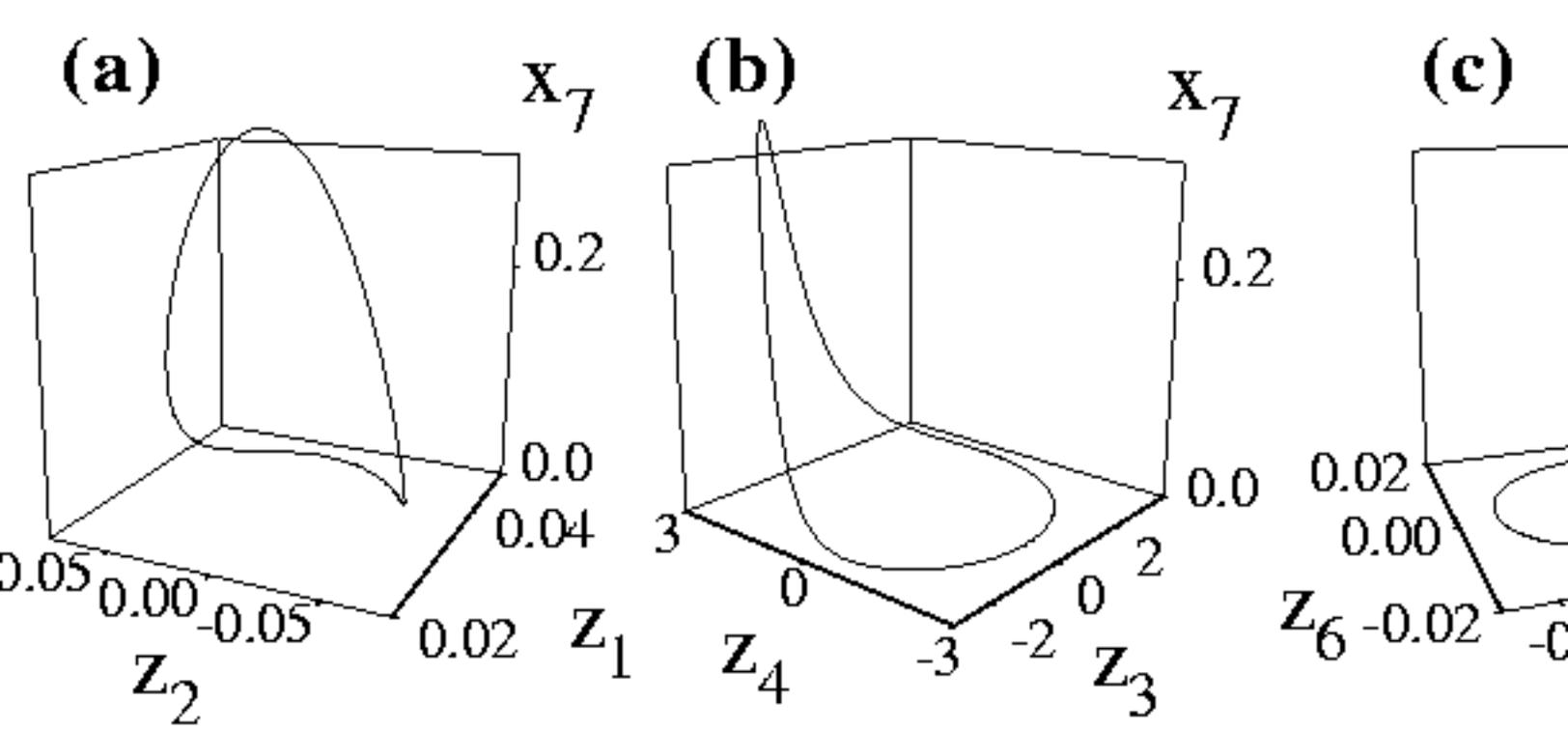
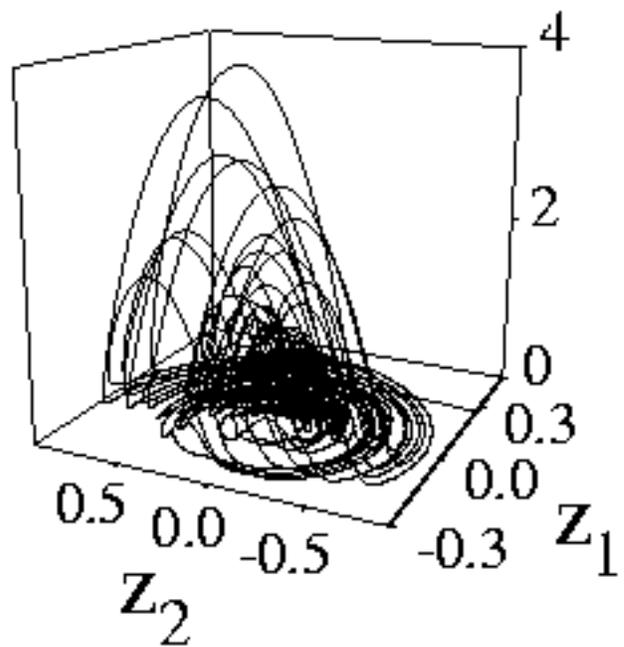
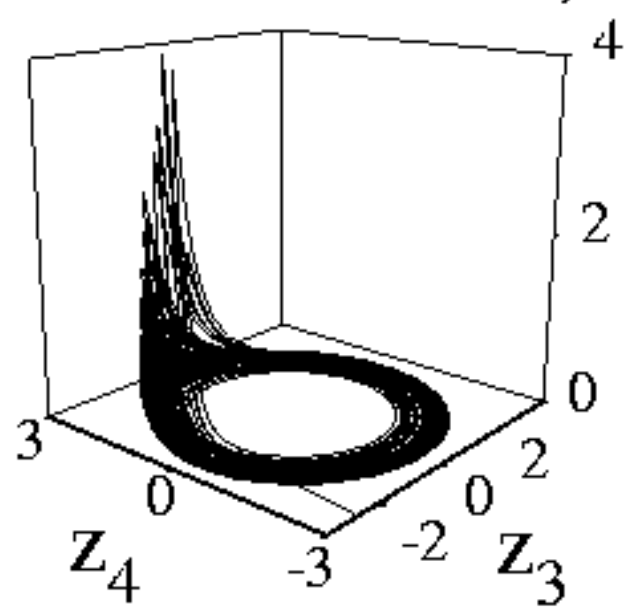
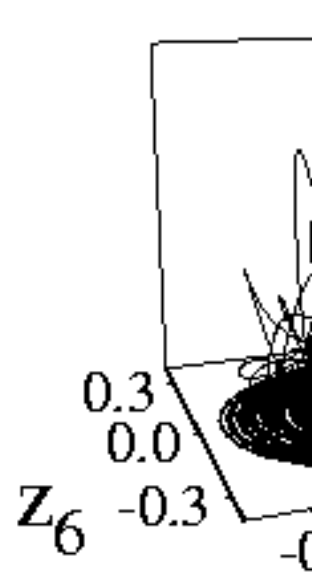
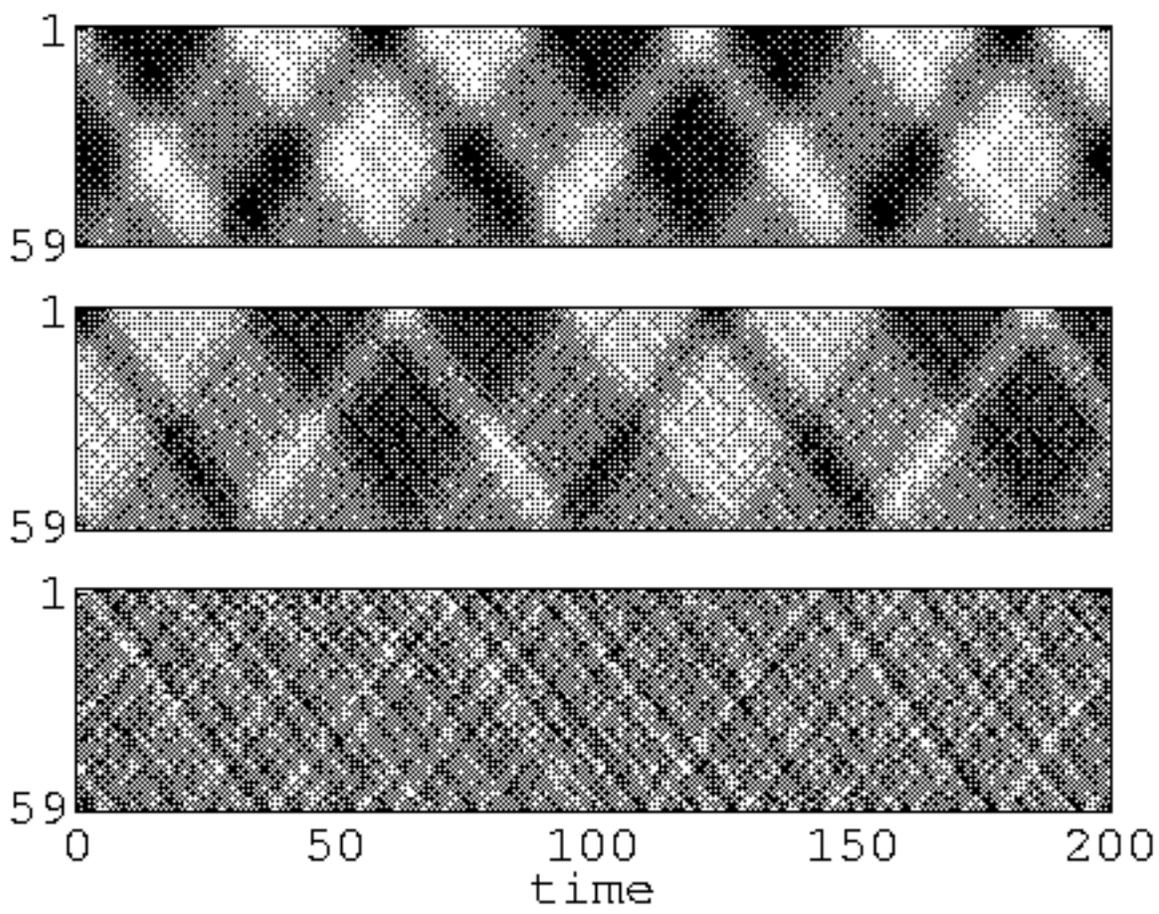
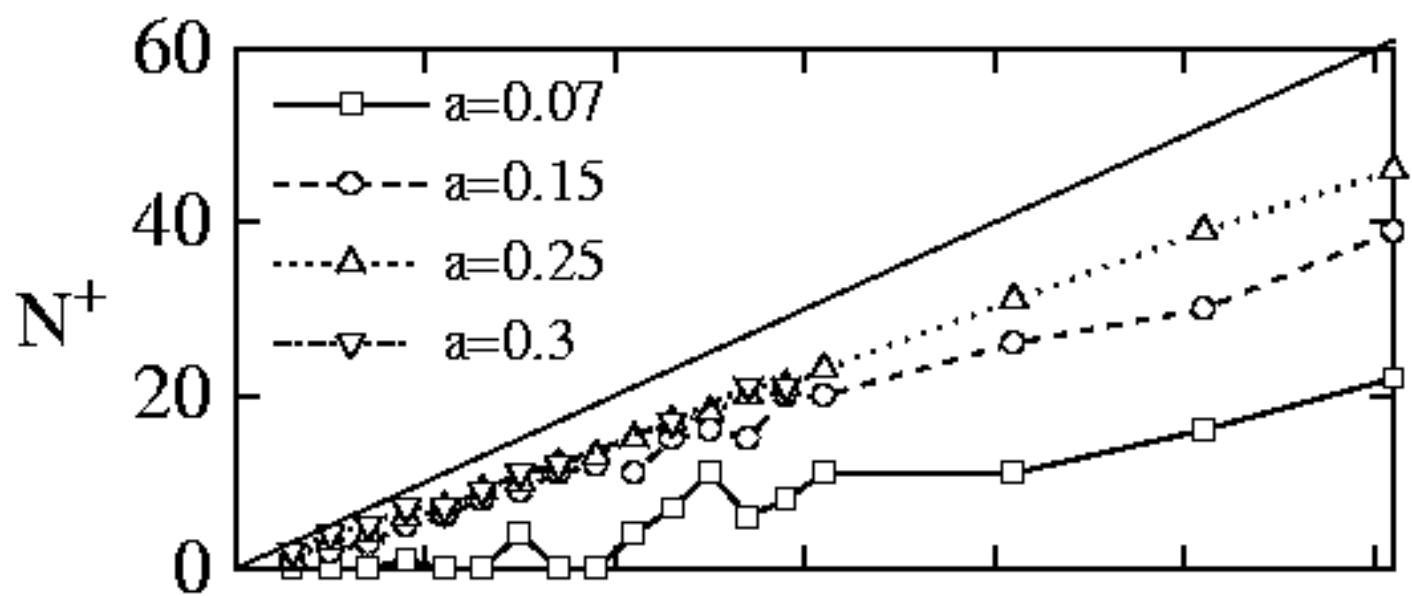
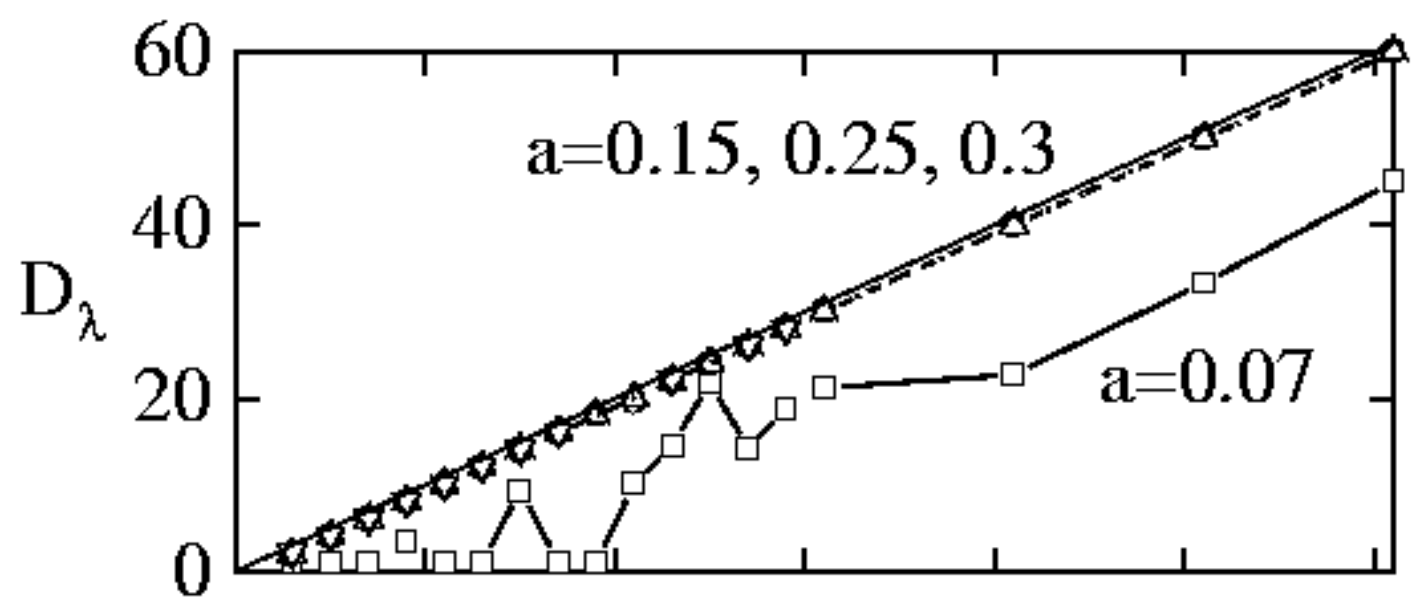
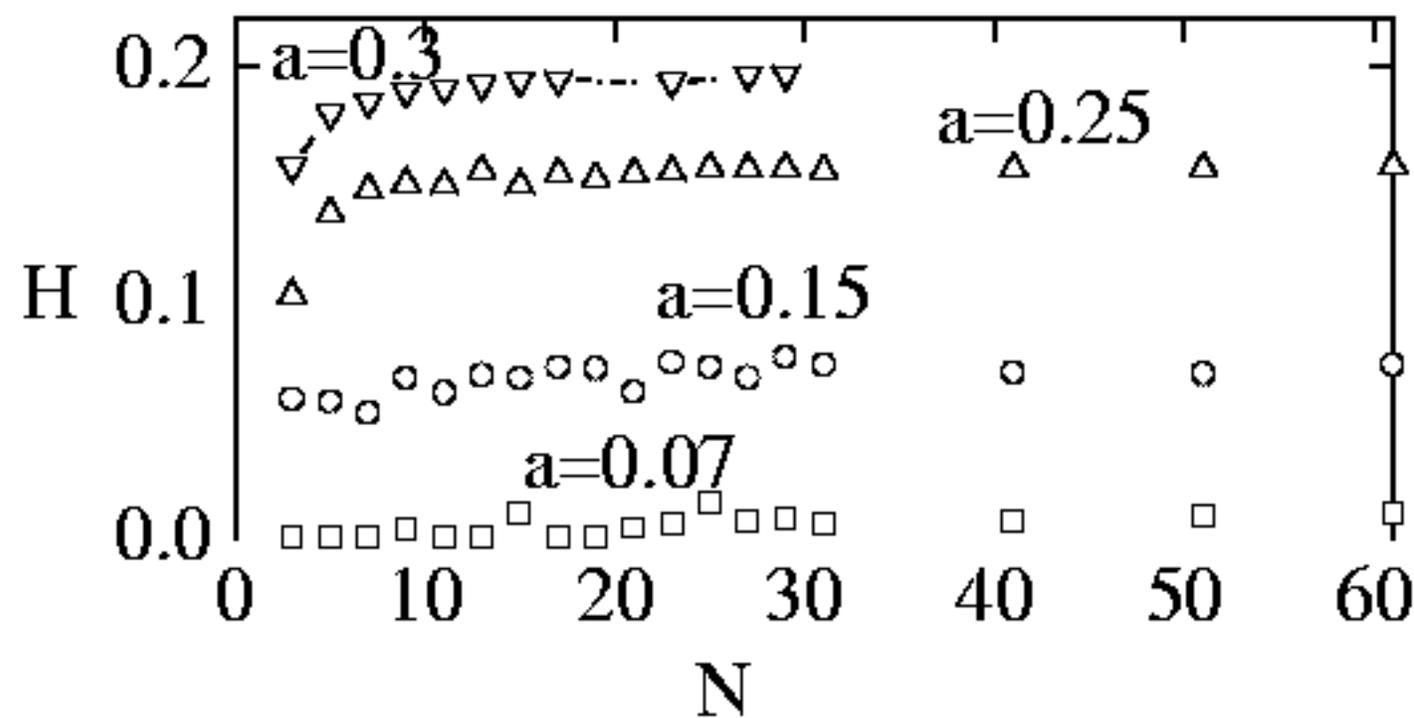
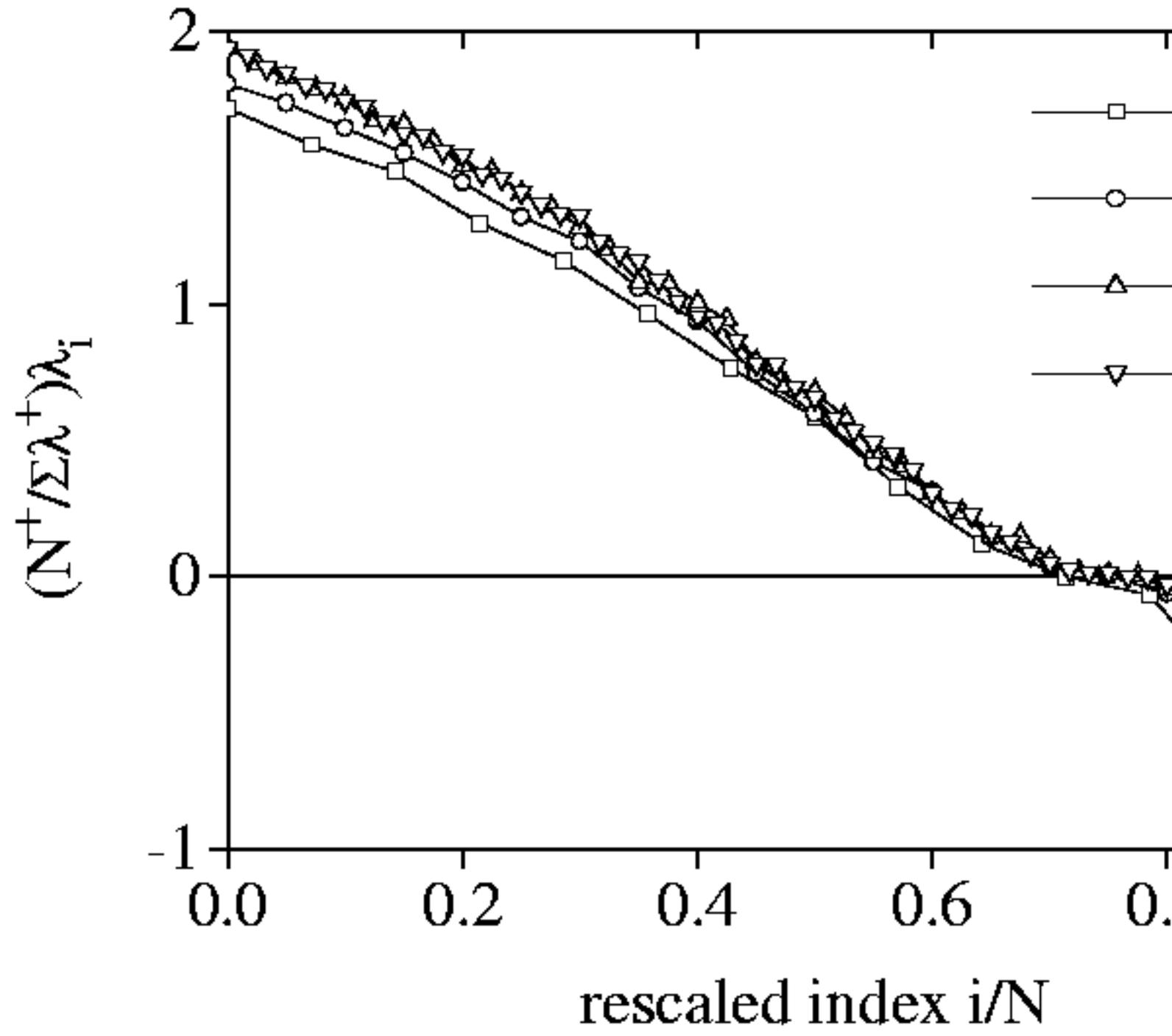
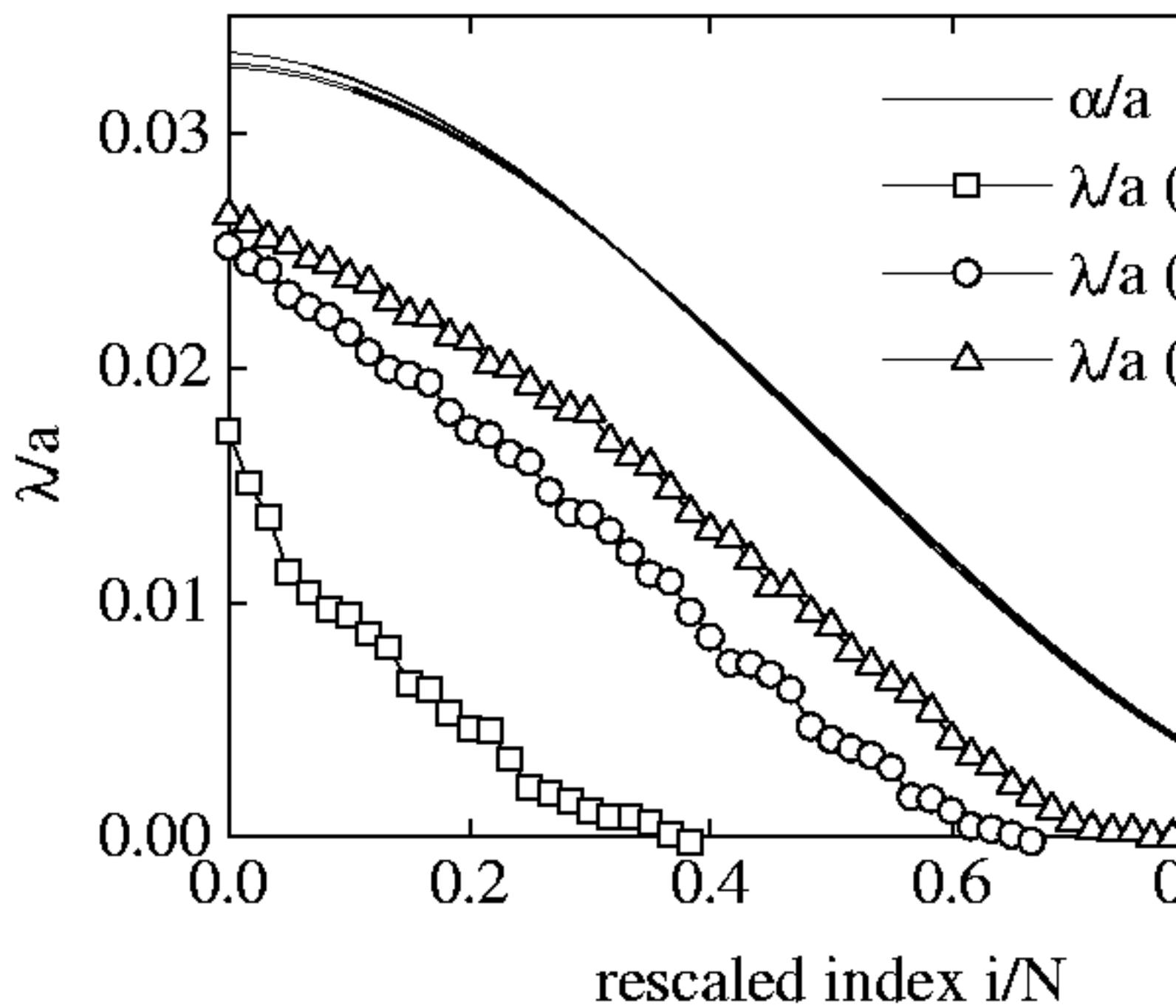
\end{document}